\documentclass[useAMS,usenatbib]{mn2e}
\usepackage{epsfig}
\usepackage{times}
\usepackage{amssymb}

\newcommand{\dd}{\mathrm{d}}
\newcommand{\msun}{M_{\odot}}
\def\apgt{\ {\raise-.5ex\hbox{$\buildrel>\over\sim$}}\ }
\def\aplt{\ {\raise-.5ex\hbox{$\buildrel<\over\sim$}}\ }
\def\degspt{$\buildrel{^\circ}\over .$}

\usepackage{graphicx}

\title[Mass estimates from stellar proper motions]
{Mass estimates from stellar proper motions: The mass of $\omega$ Centauri}
\author[R. D'Souza and  Hans-Walter Rix]
{Richard D'Souza$^{1}$\thanks{E-mail addresses: ritchiedsouza@yahoo.com
(RDS); rix@mpia.de (HWR);},
Hans-Walter Rix$^{1}$\footnotemark[1] \\
$^{1}$Max Plank Institute for Astronomy, Heidelberg, Germany\\
}

\begin{document}

\date{Accepted 2011 April 1. Received 2010 January 42; in original form
2005 January 42}

\pagerange{\pageref{firstpage}--\pageref{lastpage}} \pubyear{2010}

\maketitle

\label{firstpage}

\begin{abstract}
We lay out and apply methods to use proper motions of individual kinematic tracers 
for estimating the dynamical mass of star clusters. We first describe a simple \emph{projected mass
estimator} and then develop an approach that evaluates directly the likelihood of the
discrete kinematic data given the model predictions. Those
predictions may come from any dynamical modelling approach, and we
implement an analytic King model, a {\it spherical} isotropic Jeans equation model and an
axisymmetric, anisotropic Jeans equation model. This
{\it maximum likelihood modelling (MLM)} provides a framework
for a model-data comparison, and a resulting mass estimate, which
accounts explicitly for the discrete nature of the data for individual
stars, the varying error bars for proper motions of differing
signal-to-noise, and for data incompleteness. Each of these two methods
are evaluated for their practicality and are shown
to provide an unbiased and robust estimate of the cluster mass.
We apply these approaches to the enigmatic globular cluster omega
Centauri, combining the proper motion from van Leeuwen et al (2000) with
improved photometric cluster membership probabilities. We show that all
mass estimates based on spherical isotropic models yield
$(4.55\pm 0.1) \times 10^6 M_{\odot} [D/5.5 \pm 0.2 kpc]^3$, where our modelling allows
us to show how the statistical precision of this estimate improves as
more proper motion data of lower signal-to-noise are included. MLM
predictions, based on an anisotropic axisymmetric Jeans model, indicate for 
$\omega$ Cen that the  inclusion of anisotropies is not important for the mass estimates, but that accounting for the
flattening is: flattened models imply $(4.05\pm 0.1) \times 10^6 M_{\odot} [D/5.5 \pm 0.2 kpc]^3$, $10\%$
lower than when restricting the analysis to a spherical model. 
The best current distance estimates imply an additional uncertainty in the mass estimate of $12\%$. 
\end{abstract}


\section{Introduction}
Estimating masses of self-gravitating systems, such as
star clusters, robustly and accurately has been of long-standing 
importance in astronomy. Such
estimates must be based on the spatial distribution of some tracer
population (e.g. stars) and on its kinematics, radial velocities 
in most circumstances. However, with HST, with adaptive optics and  
the advent of the next generation telescopes, proper motion data 
have and will become more
accurate and reliable within an observation time of a few years.
Proper motions at the Milky Way's centre has already proven a boon
(e.g. \citealt{Schodel2} and references therein). 
Combined with radial velocities, proper
motions provide 5 components of the 6-dimensional
phase space for kinematic tracers. Distance estimates are usually
not accurate enough to determine the relative 6th phase-space coordinate 
\emph{within} the system. Such a wealth of
information should help to estimate accurately masses of objects
in the local group, even if accelerations are unavailable. 
Apart from the masses, such 2D or 3D kinematic data
constrain quite directly the orbit distribution and 
can uncover more subtle dynamic processes in star clusters.

Proper motion data for cluster dynamics have been first implemented
in the 70's (e.g. Cudworth 1976). However the analysis was limited by
the data quality at the time. \cite{Leo} showed that proper motion data were sufficient to
predict the dynamic mass of the cluster. However, there was no
complete proper motion data sample of a star cluster at the time.

At the present, data from HST have reached sufficient quality to
study internal cluster kinematics astrometrically 
(\citealt{McN}, \citealt{A2}). Also, diffraction limited imaging at
the VLT, Keck and other 8 m class telescopes is now a reality, and
with the next generation of telescopes coming online 
(e.g. LINC-NIRVANA camera at the LBT with
23 m resolution imaging, \citealt{He}), proper motions are
becoming a broadly competitive dynamical tool. They are not
restricted to areas with long time-baseline data, such as $\omega$
Centauri (e.g. \citealt{L1}). In addition, for many clusters, it is much 
more practical to get proper motions for very extensive data sets (1000s of stars!) 
as opposed to radial velocities. Proper motion based mass estimates scale
with the distance to the cluster as $D^3$, thereby requiring good distance estimates. Hence, it is crucial to
understand how best to use proper motion data for setting mass
limits and to learn to deal with the systematic and random errors
inherent to this technique. The main objective of this paper is to 
review and develop methods for mass estimation based on 
proper motions alone. We then apply these to determine the mass of
$\omega$ Centauri.

The Southern globular cluster $\omega$ Centauri (NGC 5139) is the most massive 
and one of the most flattened globular clusters in the galaxy (\citealt{Gey}). 
Its large mass and size sets it apart from the bulk of globular
clusters in the Galaxy. In recent years, it has evoked interest
because of the recent discovery of its multiple populations which
implies a complex formation history \citep{H2,H3}. $\omega$ Centauri
may well constitute a remnant nucleus, i.e., the former cluster
of a now disrupted galaxy (\citealt{Gne,Miz}). \cite*{Mer1}  reported a definite rotation in the cluster. 
\cite{A1} reports  that there is very little mass segregation in the cluster, 
which means that the cluster is not in equipartition.

Van de Ven et al. (2006; hereafter vdV06) have published an extensive study of 
$\omega$ Centauri using the proper motion data of \cite{L1} in addition to radial velocity data published 
by \cite{Reij} employing the Schwarzschild's orbit superposition method. 
It is important that we compare our results with vdV06, and 
seek to understand the differences that may arise. 

There has been considerable debate in recent years about $\omega$ Centauri's estimated
 distance. \cite{Long} estimates it at a distance of 5.1 kpc, while a dynamical distance of 
$4.8 \pm 0.3$ kpc has been estimated by vdV06. Using RR Lyrae stars, 
\cite{Delprincipe} estimated a photometric distance of $5.5 \pm 0.15 \pm 0.15$ 
kpc\footnote{\cite{Delprincipe} state the error correctly in distance modulus, but did not translate it correctly into distances.} 
which is in agreement with the results ($5.3 \pm 0.3$ kpc) inferred by the eclipsing 
binary OGLEGC 17 \citep{Thompson}. For this paper we adopt Del Principe's values of $5.5 \pm 0.2$ kpc.

There has also been a considerable debate on the mass of $\omega$
Centauri. Previous estimates of the mass of $\omega$ Centauri have
ranged from  $2.4 \times 10^{6} M_{\odot}$ \citep{Mandushev} to
$7.13 \times 10^{6} M_{\odot}$ \citep{Richer}. vdV06 has estimated that the mass 
of the cluster is  $(2.5 \pm 0.3) \times 10^{6} M_{\odot}$ at a dynamical distance of 4.8 kpc, 
assuming that the cluster 
is nearly-axisymmetric. A similar estimate based on an 
edge-on axisymmetric model by \cite{Mer1} sets the mass at $2.9 \times 10^{6} M_{\odot}$. 
On the other hand, \cite{Miocchi} using spherical symmetry with an anisotropic velocity 
distribution to model the cluster, gets a mass estimate of 
$(3.1 \pm 0.3) \times 10^{6} M_{\odot}$. Setting accurate
limits to the mass of $\omega$ Centauri is essential to
differentiate between the various theories of its formation
process.

Various approaches to go from kinematic data to mass estimates  have been
proposed (Binney \& Tremain 2008, BT08), differing greatly in their 
complexity. In Section \ref{sec:estimators}, we review some of the available
mass estimators and then develop better tools geared specifically 
towards proper motion data. For the present analysis, we restrict ourselves
to the self-consistent case, where the kinematic tracer density, $\nu_{*}(r)$,
is immediately related to the mass density, $\rho(r)$, that generates
the gravitational potential. This implies that the cluster is sufficiently relaxed and that 
dark matter does not appreciably change the dynamics of the cluster. In Section \ref{sec:omega-data}, 
we first describe the proper motion data set available for 
$\omega$ Centauri (\citealt{L1}, vL00 hereafter) and construct a sample of stars. 
In Section \ref{sec:omega-mass}, we apply the tools developed in this paper to the 
constructed sample to estimate the mass of the cluster. Further in Subsection \ref{sec:axijeans}, we use 
an anisotropic axisymmetric Jeans modelling technique (\citealt{Cap}) 
to account for the flattening of the cluster. We discuss the implications of our 
results and the modelling techniques we use in Subsection \ref{sec:dis}. Finally we 
summarize our results in Section \ref{sec:concl}.

\section{Mass Estimators}
\label{sec:estimators} The virial theorem has been long
and widely used as a technically simple way to estimate the masses 
of self-gravitational systems from the motions of individual tracers 
or constituents.
However, a major drawback of estimators that are based on the
virial theorem is that they tend to be  biased and formally have
an infinite variance. That led e.g. \cite{H1} to develop a practical 
``projected mass estimator'' for self-gravitating systems.

While for many systems (e.g. extra-galactic ones),
radial velocities are the only observable kinematic component,
\cite{Leo} extended the estimator to include proper motions, and applied this
to estimate the mass of the open cluster M35.

In this Section, we shall outline two estimators that have many conceptual
and practical improvements. We then see how they stack up against the Jeans 
equation modelling, which exploits more information (e.g. \citealt{Wolf}, for a
recent application).

\subsection{Projected Mass Estimator (PME)}
\label{sec:PME}Projected mass estimators try to exploit the directly 
observable, projected properties of a self-gravitating system, while being
efficient and unbiased in obtaining mass estimates of finite variance.
In practise, such estimators have been derived by taking moments of the Jeans equation for 
a spherical system. For example, a projected mass estimator for proper motion 
velocities can be written as
\begin{equation}
M = \frac{32}{3 \pi G} <(2 v^{2}_{R} + v^{2}_{T}) R>,
\label{eqn:proj1}
\end{equation}
where $v_{R}$ and $v_{T}$ are the radial and tangential proper
motion velocities, $R$ is the projected distance and $M$ is the
total mass of the cluster (\citealt{Leo} Eq. 19).

Note that such a projected mass estimator uses both components of the 
velocity dispersion, thereby directly  accounting for any anisotropy in the system.

Since any projected mass estimator involves a line-of-sight
projection and radius averaging, care must be taken to have a proper radial sampling of
the tracers of the system. In this estimator, particles at larger 
projected distances are given more
weight than particles at smaller projected distances, because of
the inability to reconstruct the true radial distance for
particles at small projected distance.

Such projected mass estimators become biased when the velocity
measurements have significant errors, $\delta v$. The resulting mass over-estimate
grows quadratically with the velocity error. However, we
can modify the projected mass estimator to take into account the
velocity errors. Since $ <v^{2}_{obs}>  =  <v^{2}_{true}> +
<\delta v^{2}>$, a reasonable unbiased estimator would be
\begin{equation}
M = \frac{32}{3 \pi G} <(2 v^{2}_{R} + v^{2}_{T} - \delta v^{2})
R>, \label{eqn:proj2}
\end{equation}
where $\delta v^{2} = 2 \delta v_{R}^{2} + \delta v_{T}^{2}$.

Note that such mass estimators do not provide a ``formula'' for estimating the 
error in the inferred mass. However, bootstrapping methods can be used to obtain
such error estimates.

Correcting statistically for the velocity errors as in the above
estimator is a statistical process, and may fail for a small 
number of particles. Any over- or under-estimation of the
proper motion errors would again bias any estimate.

\begin{figure*}
  \includegraphics[width = 80mm]{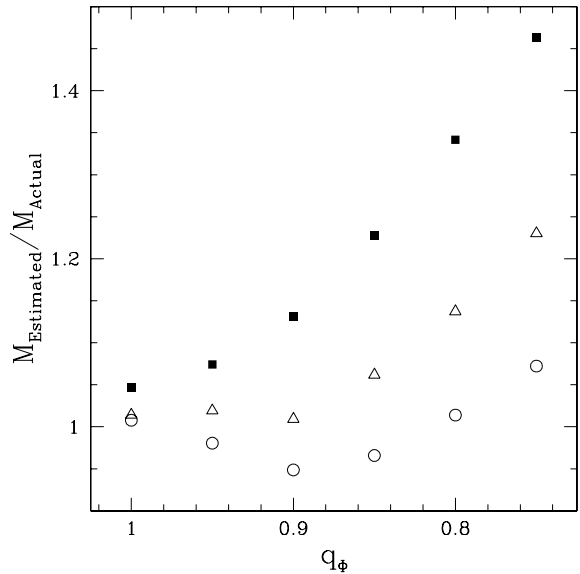}
  \caption{The role of flattening in mass estimates: the figure shows the ratio between estimated and true mass values 
      using the PME  for the N-Body realization of a lowered Evans model cluster (Kuijken \& Dubinski 1994) 
      as a function of the flattening parameter of the potential $q_{\Phi}$ at various levels of inclination. 
      The filled squares are the ratio between estimated and true mass values for a face-on view of the cluster 
      (an inclination of $0^\circ$). 
      The open triangles and the open circles give the ratio between estimated and true mass values when the 
      cluster is inclined at an angle 
      of $50^\circ$ and $90^\circ$ (edge-on) respectively. Note that flattening in the potential of $q_{\Phi} 
      \approx 0.9$ corresponds to stronger flattening in the density $q_{\rho} \approx 0.7$.
  }
  \label{fig:mass}
\end{figure*}

Note also, that the simple  projected mass estimator and the virial
mass estimator assume radial symmetry. If the cluster shows deviation 
from radial symmetry or is flattened, the mass estimates will be biased.
We can quantify how much the projected mass estimator is biased with 
increasing flattening by constructing N-Body realizations of lowered 
Evans model \citep{Ku} with decreasing potential flattening ($q_{\Phi}$). 
We then apply the projected mass estimator (Eq. \ref{eqn:proj1}) to synthetic
error-free data from a face-on view: the bias in the resulting (spherical) mass
estimate depends on the flattening and on the inclination of 
the cluster as seen in Figure \ref{fig:mass}. Given that $q_{\Phi} \approx 0.9$ 
corresponds to a three times flatter mass distribution, the bias is modest ($10\%$).

It is important to emphasise that the PME is an averaging process 
across the entire radial profile of the cluster. This poses practical 
challenges both at the cluster centre, due to image crowding and 
at the outer edge of the cluster due to contamination from physically
unrelated stars. As \cite{Genz} and \cite{Schodel2} have pointed out, moment
estimators must be applied cautiously in cases where the data does not extend 
over the entire length of the cluster, or where the observed sample is 
dominated by stars that are intrinsically far from (but projected near to) 
the centre. 

To summarize, the ``projected mass estimator'' is very easy to implement, 
but has a number of practical and conceptual limitations. Therefore, we now 
turn to explore more (spatial) information.

\subsection{Maximum Likelihood Models (MLM)}
\label{sec:MLM} Here we outline an approach to estimate cluster masses that is much
more geared towards the discrete data that is actually available: individual proper motion
measurements, with errors that are significant w.r.t. the velocity dispersion. Good 
estimators should use high-precision measurements effectively and  not overestimate the 
information content of low-precision velocity measurements. This can be accomplished
in a framework where we directly calculate the likelihood of the data given some model 
predictions, and then maximize it w.r.t. different models (e.g. differing in mass).
Such an approach must be based on families of suitable models that allow to predict the 
components of the proper motion velocity dispersions (both radial and tangential) 
as a function of the projected radius ($R$). This can be done by either drawing on
an analytic model (e.g. \citealt{K66}), or on an orbit based model (vdV06), 
or on the Jeans equation. If assumptions are made about the symmetry (ellipticity)
and anisotropy of the cluster, the models reduce to a one parameter family, and the data
then simply set the mass normalization. These model approaches are well established;
what is new here is the direct data likelihood estimate.

Let us presume that we can model the projected proper motion velocity
distribution at each radius as a Maxwellian distribution. Thus the
probability of a star in the cluster to have a certain proper
motion velocity $v$ at the projected distance $R$ in the centre-of-mass 
frame assuming no rotation given that the total mass of the cluster is
$M$ is:
\begin{equation}
p_{model}( v | M ) =  \frac{1}{\sqrt{2 \pi \sigma^{2}(M,R)}} \exp
\left\{ \frac{- v^{2}}{2\ \sigma^{2}(M, R)} \right\}, 
\label{eqn:model}
\end{equation}
where $\sigma$ is the projected velocity dispersion which scales with the
total mass of the cluster as $\sigma(M, R) = \sqrt{M} \sigma(R)$.  

On the other hand, every observation of proper motion velocity
$v_{i}$ has a certain  error $\delta v_{i}$ associated with it.
The probability of actually measuring the velocity $v_{i}$ at a
projected distance $R_{i}$ is a Gaussian distribution given by:
\begin{equation}
p_{obs}(v_{i} | \, v, \delta v_{i} ) = \frac{1}{\sqrt{2 \pi\
    \delta v^{2}_{i}}} \exp  \left\{ \frac{- (v-v_{i})^{2}}{2\ \delta v^{2}_{i}}  \right\}.
\label{eqn:pobs}
\end{equation}

Thus for a cluster of mass $M$, the likelihood of measuring a proper motion velocity
$v_{i}$, at a projected distance $R_{i}$ with an observational error $\delta v_{i}$ is,
\begin{equation}
p_{i}(v_{i} | \, M) = \int p_{obs}(v_{i}| \, v , \delta v_{i}) \, p_{model} (v | \, M)\, \mathrm{d}v.
\label{eqn:convol}
\end{equation}

Note that the convolution in Eq. \ref{eqn:convol} `down weighs' the impact 
of the data with large errors in discriminating models of different mass.

The log-likelihood of the cluster to have a mass $M$ given $N$
observations is:
\begin{equation}
\mathcal{L}(\forall  v_{i} |\, M) = \sum_{i=1}^{N}\ln p_{i}(v_{i} | \, M),
\label{eqn:like}
\end{equation}
where the sum is over both radial and tangential proper motion
velocities.

The corresponding (posterior) probability distribution for the model parameter $M$
is maximised for $\mathcal{L}_{max}$ of the likelihood function. 
Confidence limits ($1\sigma$)
are given by the interval for which $\mathcal{L}_{max} - \mathcal{L} = 0.5$ 
(c.f. \citealt{Lamp}, whose statistic S $\equiv 2 \mathcal{L}$; \citealt{Efs}).

Note that the maximum likelihood methods do not directly involve the density profile of the tracers, $\nu_{*}(r)$, in the calculation of the mass of the cluster. Hence the observable shape and characteristics of the tracer density profile, $\nu_{*}(r)$ need not correspond to the actual mass density profile of the cluster, $\rho(r)$. Rather the maximum likelihood estimator only asks what the proper motions of individual stars are \emph{given that we have this measurement at radius} $R_{i}$ and a certain dynamic model.
 
The maximum likelihood estimator can be extended to axisymmetric models, if the axisymmetric Jeans
equations are used to model the velocity dispersions and the rotation of the cluster \citep{Bin,Mar,Ems,Cap}. 
Equation \ref{eqn:model} can be suitable modified as:
\begin{equation}
p_{model}( v | M ) = \frac{1}{\sqrt{2 \pi\ \sigma^{2}(M, X, Y)}} \exp
\left\{ \frac{- (v - \overline{v}(M, X, Y)) ^{2}}{2\ \sigma^{2}(M, X, Y)} \right\}
\label{eqn:rotation}
\end{equation}
where the mean velocity, $\overline{v}$ and the velocity dispersions depend on the projected position $(X,Y)$ and scale with mass appropriately. 

\cite{Cap} has proposed a convenient way to model the anisotropic axisymmetric Jeans equation by extending the Multi-Gaussian Expansion (MGE) method of \cite{Ems}. We derive the necessary formulas for the proper-motion velocity dispersion and the rotation in Appendix A.

\section{$\omega$ Centauri Data Sets}
\label{sec:omega-data} In this Section, we discuss the available data sets on $\omega$ 
Centauri, and the pre-selections and pre-processing that is needed to implement the
concepts of Section \ref{sec:estimators}.

\begin{figure*}
   \includegraphics[width=0.33 \textwidth]{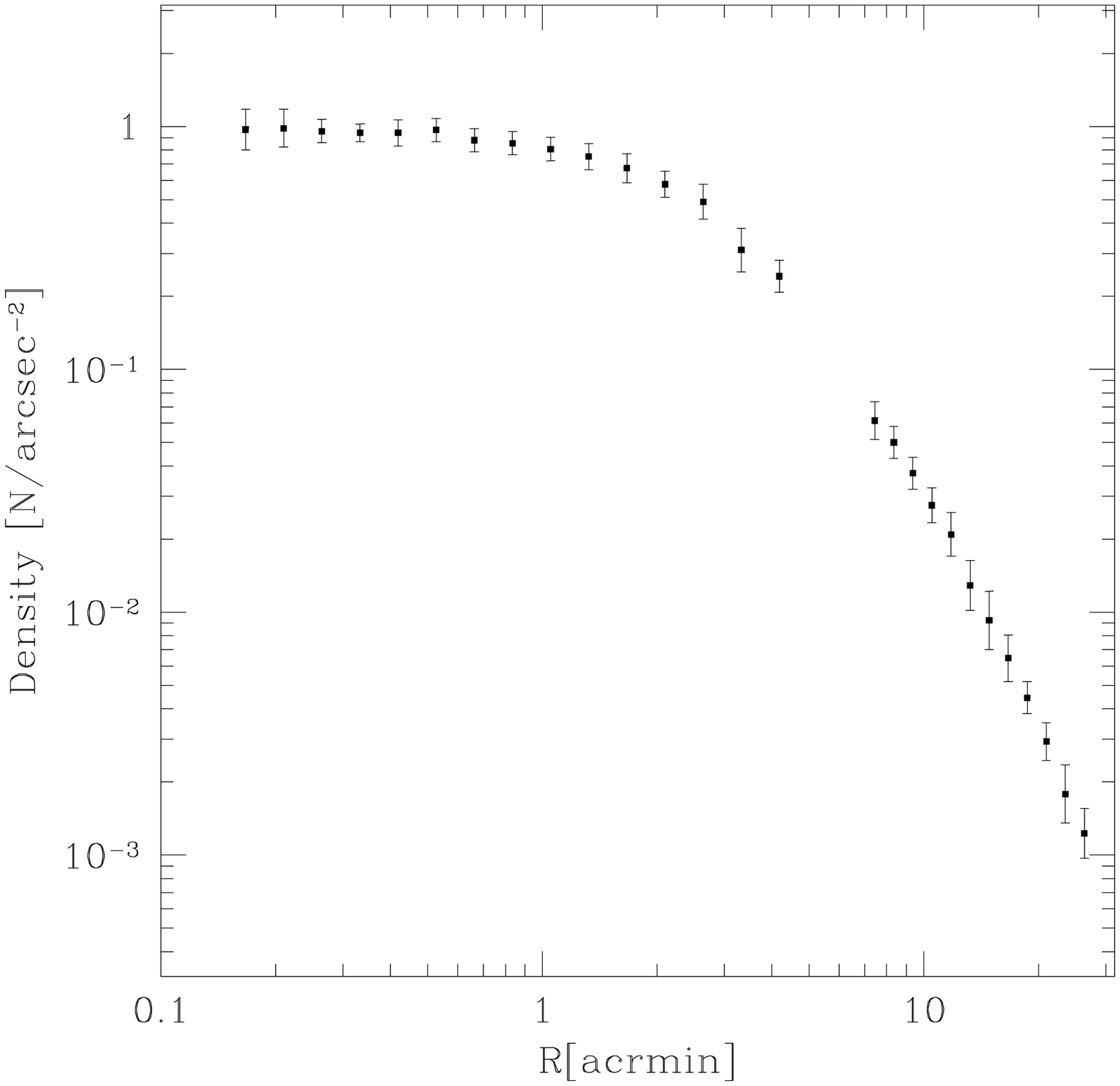}
   \includegraphics[width=0.33 \textwidth]{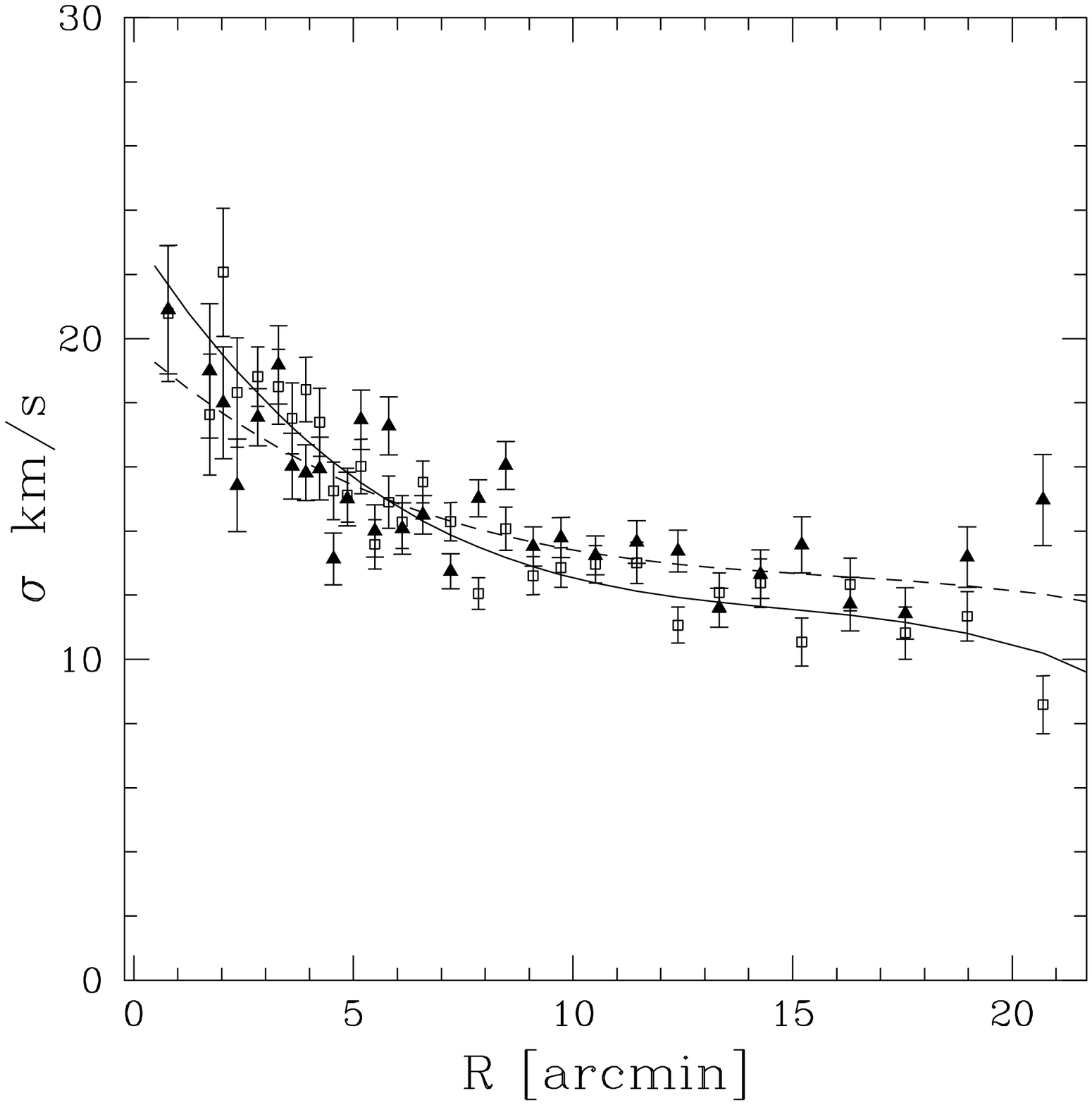}
   \includegraphics[width=0.33 \textwidth]{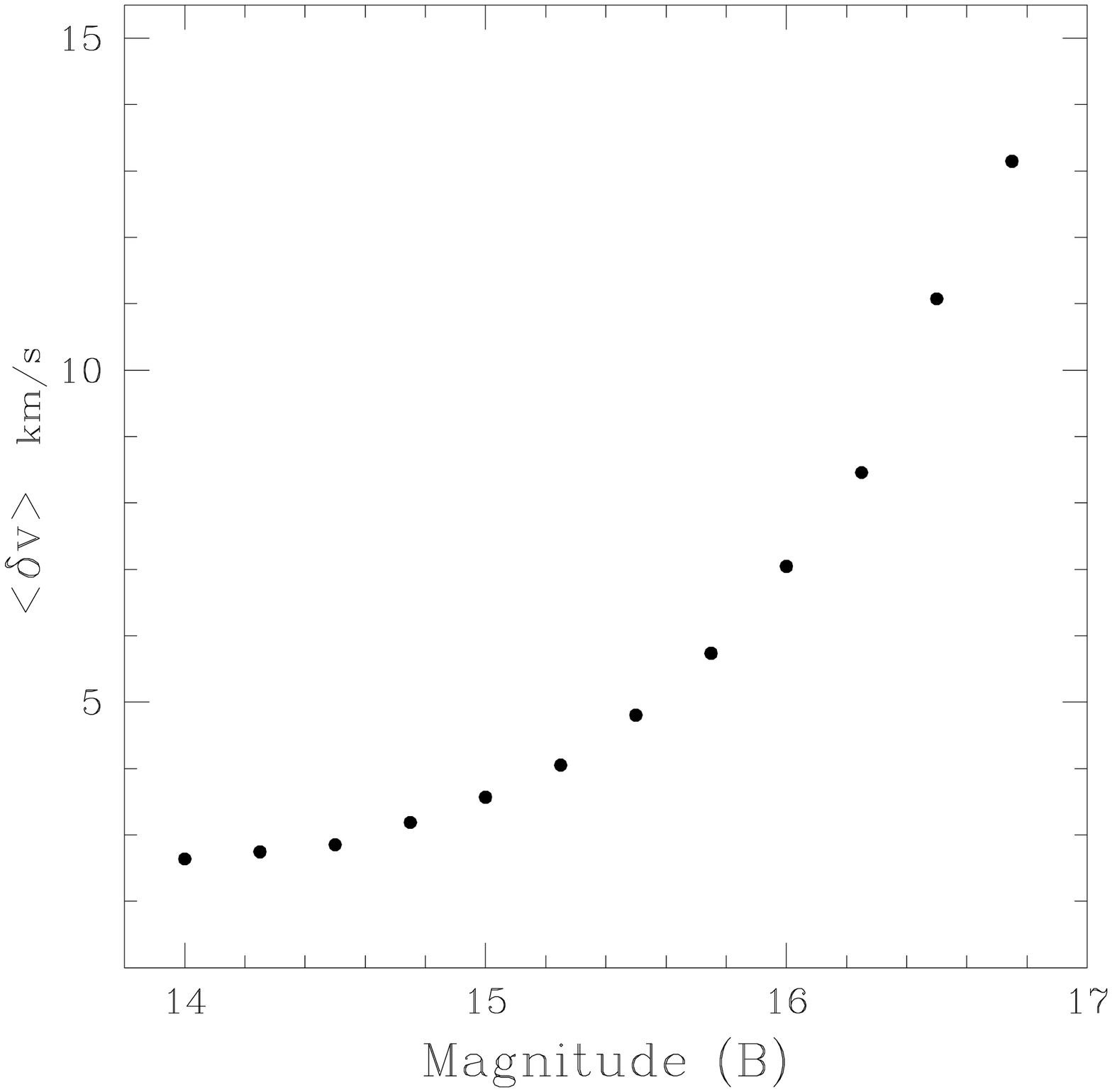}
   \caption{ Observational constraints for proper motion based mass 
	estimates of $\omega$ Centauri : (a) \emph{Left}: the normalized surface number density profile of \citet{Fer}
	in logarithmic scale. (b) \emph{Centre}: radial (squares) 
	and tangential (triangles) proper motion dispersion profiles as a 
	function of projected radius of Sample C at a distance of 5.5 kpc (see Section \ref{sec:dataset}). 
	The solid and dashed lines represent splines fit through the radial and tangential velocity 
	dispersions respectively. (c) \emph{Right}: typical proper-motion errors of
	Sample C (see Section 3.2) stars used in Section 4 to determine the mass of the cluster as a function
	of apparent magnitude.}
   \label{fig:problem}
\end{figure*}

\subsection{Photometric Data Sets}
No two dimensional surface brightness profile of $\omega$ Centauri have been published so far.
The V-band surface brightness data of \cite{Mey87} has been used most in the 
literature. Recently \cite{Fer} have published a more accurate surface density profile  
which differs from \cite{Mey87} in its outer parts. For this paper, we use the number density profile of \cite{Fer}, which 
has been reproduced in Figure \ref{fig:problem}. $\omega$ Centauri is significantly flattened 
\citep{Gey, White}. The cluster has a tidal radius $= 45'$ \citep{Trager}. 

\cite{Rey2} have published wide-field and high precision \emph{BV} photometry in the field of
$\omega$ Centauri for a large number of stars for $V > 22$. Using Ca and St\"omgren \emph{by} photometry \citep{Rey1}, 
they separate the field stars (for $V > 16$) using the metallicity-sensitive \emph{hk} index defined as
(Ca - \emph{b}) - (\emph{b} - \emph{y}) plotted against (\emph{b - y}). Because even the most
metal-rich stars in $\omega$ Centauri are relatively metal-deficient compared with the 
typical disk field stars, \cite{Rey2} were able to eliminate foreground field stars through 
the use of the \emph{hk} index. We use this information below to assign photometric 
membership probabilities.

\subsection{Proper Motions}
vL00  have provided a large set of proper motion data
for $\omega$ Centauri drawing on archival photographic 
image plates. These proper motions are based on 100 plates
obtained with the Yale-Columbia 66 cm refractor, covering the epochs 1931-1935 and 
1978-1983. Differential proper motions were obtained for 9847 stars for a limiting photographic magnitude of 
$16.0$ for the centre of the cluster and $16.5$ for the outer parts. The precisions of the proper motions range from $\sim 0.1$ mas/yr 
for the brightest to an average of $\sim 0.65$ mas/yr 
for the faintest stars. These data have already been used previously (vL00, vdV06)
to study the internal kinematics of the cluster.

Other proper motion data sets also exist. Recently, \cite{Bel} have made available 
the first ground-based CCD proper motion catalogue of $\omega$ Centauri with a 4 year 
baseline using a Wide Field Imager at ESO's 2.2 m telescope. For stars within their saturation limit 
($V > 14.6$) to the vL00 faint limit ($V \sim 16.5$), the estimated error is $\sim 0.75$ mas/yr. 
\cite{A2} have published high-quality proper motions using the HST's ACS
for 53,382 stars in a central $R \lesssim 2'$ field limited to stars on or below the 
sub-giant branch and those brighter than $m_{F435W} = -11$, a few magnitudes below the turnoff, 
using a 2.5- to 4- year baseline. Their typical proper-motion error is better than $0.1$ mas/yr.


As we are interested in the total mass of $\omega$ Centauri (within its tidal radius), 
we focus on the large-area, ground-based data, rather than HST data (e.g. \citealt{A2}).
For our mass estimates of $\omega$ Centauri, we restrict ourselves to  the proper motion data of
vL00.

As shown in Figure \ref{fig:problem}, the errors in the proper
motion data becomes comparable to the velocity dispersion profile
at magnitude 16, especially for stars in the outer parts of
the cluster. Thus reasonable and robust estimates of the mass can
be obtained at around magnitude 16.

\subsection{Sample Selection and Pre-processing of Data for the Mass Estimates}
\label{sec:dataset}
\begin{figure*}
  \includegraphics[width=0.49 \textwidth]{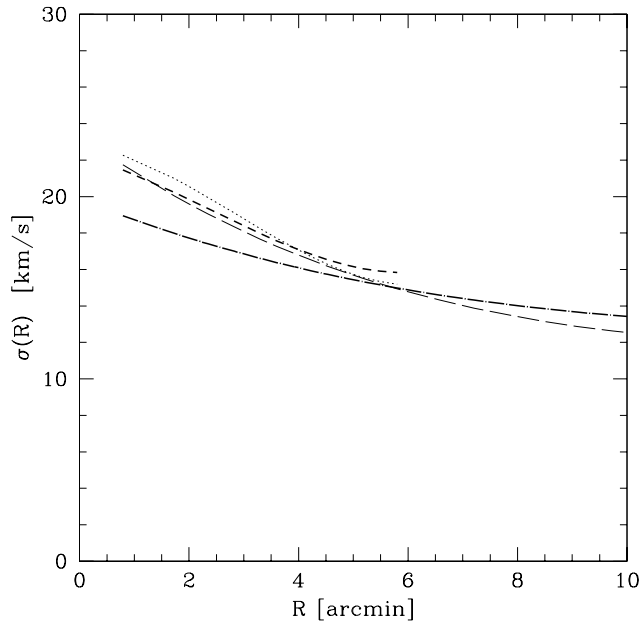}
  \caption{The projected radial (long dashed line) and transverse (dark dot long dashed line) velocity dispersion of 
	stars of Sample C (as defined in Section \ref{sec:dataset})  of class 0 stars only 
	and the projected radial (dotted line) and transverse (dark short dashed line) velocity dispersion of sample of class 1 stars 
        only as a function of radial distance. Sample C is defined later in the text. The sample of class 1 stars was chosen in 
        a similar way to Sample C, but using only class 1 stars. The velocity dispersion of the sample of class 1 stars extends radially upto $6'$. 
	Class 1 stars have a much higher tangetial velocity dispersions (dark short dashed line) 
	in the central part of the cluster than class 0 stars (dark dot long dashed line).
	}
   \label{fig:class_comp}
\end{figure*}

vL00 divide their proper motion observations into 4 classes
based on the errors  in the proper motion determination which
depend upon crowding and plate quality. We only use stars with well measured 
proper motions (class 0) since their errors appear to be relatively well behaved (see Fig. 11 of vL00).
Class 1 stars are slightly disturbed by a neighbour and are localized in the central part of the cluster.
A comparision of the velocity dispersion curves in Figure \ref{fig:class_comp} indicates that class 1 stars 
have a higher tangential velocity that class 0 stars in the central $5'$ of the cluster.

\subsubsection{Pre-processing of the Data}
Following vL00, the centre of the cluster was taken to be $\alpha
= $ 201\degspt69065, $\delta= $ -47\degspt47855. To start off, we assume that the
distance of the cluster is 5.5 kpc.
We project the absolute celestial coordinates $\alpha$ and $\delta$ into
Cartesian coordinates (relative to the cluster centre) onto 
the plane of the sky perpendicular to the line-of-sight vector through the 
cluster centre \citep{Ko}:
\begin{eqnarray}
x & = & - f_{0} d \cos\delta\sin\Delta\alpha \nonumber \\
y & = & f_{0} d (\sin\delta\cos\delta_{0} - \cos\delta\sin\delta_{0}\cos\Delta\alpha)
\end{eqnarray}
and
\begin{eqnarray}
d & = & (\sin\delta\sin\delta_{0} + \cos\delta\cos\delta_{0}\cos\Delta\alpha)^{-1}
\end{eqnarray}
where $f_{0}$ is a scaling factor ($f_{0} = 1$ to have $x$ and $y$ in radians
and $f_{0}= 180/\pi$ to have $x$ and $y$ in units of degrees), $\alpha_{0}$
and $\delta_{0}$ are the cluster centre coordinates, and $\Delta \alpha \equiv \alpha
- \alpha_{0}$ and $\Delta \delta \equiv \delta - \delta_{0}$.

We corrected the proper motion data for apparent rotation arising purely from 
projection effects using a canonical distance of 5.5 kpc 
and a systematic l.o.s velocity of $232.02$ km/s (vdV06). 
Any solid-body rotation cannot be determined from the differential proper motions
due to the astrometric reduction process employed to measure the proper motions in the first place.
However it can contribute significantly to the kinematics of the cluster. We correct for 
such a  residual solid-body rotation with $\Omega = 0.029$ mas/yr/arcmin following vdV06.

\subsubsection{Construction of Samples}
\begin{figure*}
  \includegraphics[width = 0.49 \textwidth]{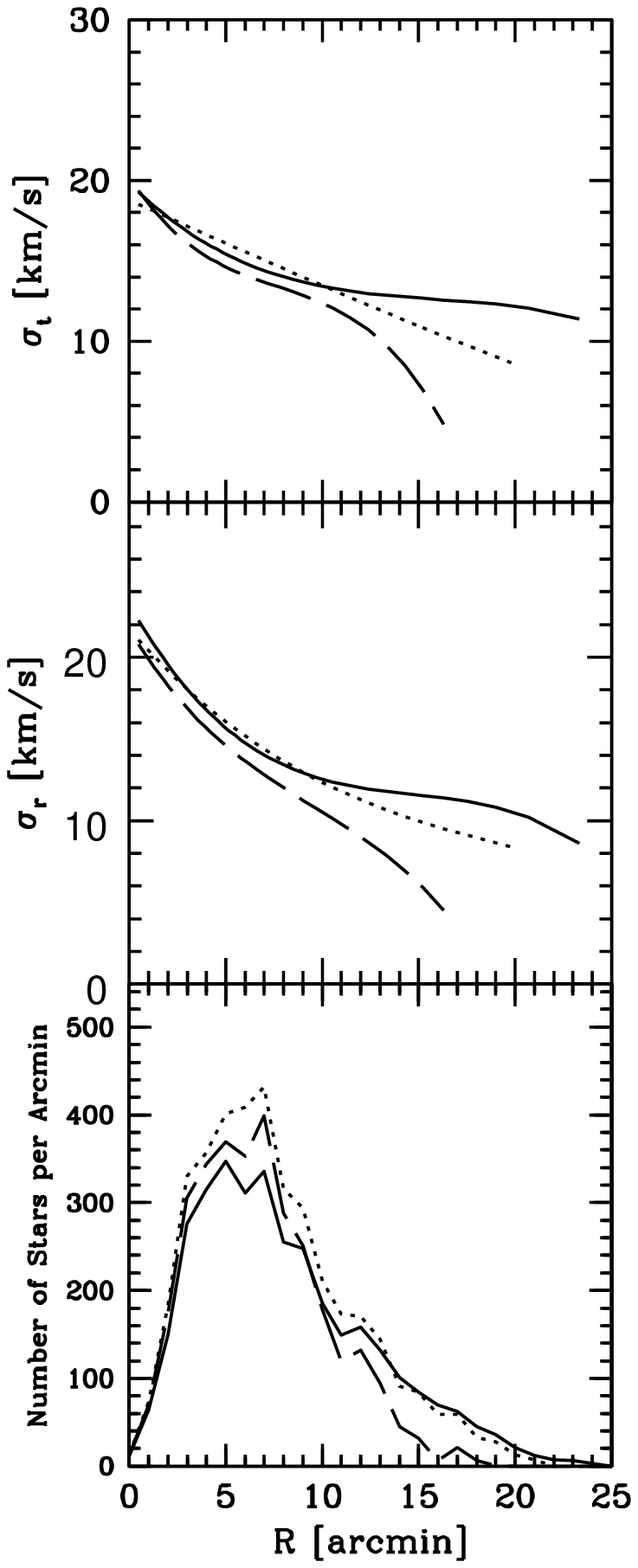}
  \caption{The number of stars per radial bin and the radial and tangential velocity dispersions of three samples
	as a function of projected radial distance. The thick line represents a predominately photometrically 
	determined Sample C (3384 stars). The long-dashed line represent a sample of stars whose membership probability is greater than
	$98.5\% (2.5 \sigma)$ (3881 stars). The short-dashed line represent a sample of stars whose
	membership probability is greater than $99.3\% (3 \sigma)$ (3201 stars). 
	The estimated mass of the cluster also varies   
	accordingly as $4.58 \pm 0.09$, $4.48 \pm 0.08$ and $3.62 \pm 0.06 M_{\odot} \left[D/5.5 \pm 0.2 kpc\right]^{3}$
	respectively. The mass is derived using a MLM of a single isotropic Jeans model.
  }
  \label{fig:sample_comparision}
\end{figure*}

vL00 calculate the membership probability of each star on the basis of its velocity and 
its projected distance. Choosing cluster members based merely on these membership probabilities 
is fraught with error, as they are correlated with radial completeness, mean radial distance 
and the limiting magnitude. Figure \ref{fig:sample_comparision} demonstrates how the completeness
and velocity dispersion of the sample correlates with the assigned membership probability, thereby
affecting the mass estimate. Samples with higher cut-off membership probabilities are not only radially 
incomplete but also have lower velocity dispersion curves and hence lower estimated masses. 

On the other hand, vdV06 select their sample by choosing those stars whose proper motion 
errors were lower than $0.20$ mas/yr after a proper motion membership cutoff of at least 68 percent, 
obtaining a limited sample of 2295 stars. 

For the purpose of kinematic studies it is crucial to determine cluster membership in a way that 
is independent of the star's velocity. In the subsequent analysis, we will therefore base cluster membership 
predominately on photometric properties. 

In the trade-off between purity and size of photometrically selected samples, we consider two cases. 
\emph{Sample A}: we consider only the horizontal branch stars in vL00 data, where field contamination
is apparently negligible. Selecting horizontal branch stars in the ranges $-0.1 < B-V < 0.4$ and $14.4 < V < 16$, 
we get a total of 1053 stars.

\emph{Sample B}: we correlate the \cite{Rey2} published  sample of cluster members based on photometry with the 
vL00 data (Sample B). The high precision \emph{BV} photometry of \cite{Rey2} data combined with the 
Ca and St\"omgren \emph{by} photometry \citep{Rey1}, efficiently separates cluster stars from the
typical disk field stars. To match the vL00 and \cite{Rey2} data-sets, we first had to align them. 
We mapped the two data-sets using the IRAF CCMAP routine. We then cross-matched the two 
data-sets in a four vector space (RA, DEC, B and B-V). For the magnitude and colour, we gave 
a large leeway of 0.5 and 0.1 respectively. The size of the search window of the RA-DEC was 
$0.0008^\circ$ ($\sim 3\arcsec$). We matched a total of 3321 stars.

However, this photometric approach leads to manifest outliers, stars
that photometricaly look like $\omega$ Centauri members but whose velocity is far greater than the 
presumed escape velocity. Such stars will bias any mass estimates. Therefore, it is important that we
also clean the sample kinematically of such outliers while introducing as little bias as possible. To do 
this, we calculate iteratively the velocity dispersion and excise stars outside $\pm 3 \sigma$, 
until all the stars are within the dynamical bounds. For Sample A, this clipping removes $<10\%$ 
of the stars leading to 968 members. For Sample B, $2.4\%$ stars are removed by the $3 \sigma$ clipping, 
bringing the numbers of stars in the sample to 3241.

Using the radial velocity data of \cite{Reij}, we can derive an independent rough estimate of the number 
of field stars in our sample. For the $3 \sigma$ cleaned Sample B, out of the total of 1028 matched stars 
between the two samples, 22 stars were found to be field stars. The probability of the number of fields 
stars in the sample is $2.14\%$. Given the additional data made available by the radial velocity data, we
can do better.

\emph{Sample C}: both Sample A and Sample B suffer from radial incompleteness, with Sample B 
extending partially upto R$=22'$.
We can extend the sample with the help of the radial velocity 
data of 1589 cluster stars of $\omega$ Centauri published by \cite{Reij}. By using only stars of class 0, we 
have 1276 stars which are 100\% kinematically-certified cluster members that extend from $3'$ to $30'$. Removing 
the 1028 common stars and the 22 field stars of Sample B, we get a total of 3467 stars.
After a 3-sigma clipping, we get 3384 stars (Sample C). Out of these 3335 stars are within 20'. 

Following vdV06, we only use stars within $20'$ for the ML estimators. The ML estimators are independent of the 
spatial completness of the sample. Possible field stars with high velocities $>20'$ may unnecessarily bias the ML estimator. 

On the other hand, we use the complete Sample C for the PME so as not to bias the estimator at the outer edges of the cluster. In the PME, the estimated mass involves an averaging process over the entire length of the cluster and hence completness is important. However, any sample so constructed is still incomplete near the center of the cluster due to crowding effects. This issue was raised in Subsection \ref{sec:PME} and will be partially addressed in Subsection \ref{sec:radprof}.

Therefore in the final analysis we will only use Sample C stars with the ML estimators for the cluster mass estimates. Sample A stars are limited in number and have large errors associated with their proper motion. Similarly Sample C is superior to Sample B in that it has a slightly higher number of stars and fewer contaminated field-stars. Nevertheless, we will also implement the PME with Sample A and Sample C in Subsection \ref{sec:omega-mass-pme} to get a feel as to how the estimator fares.

\subsubsection{Velocity Dispersion}
\begin{figure*}
  \includegraphics[width= 0.48 \textwidth]{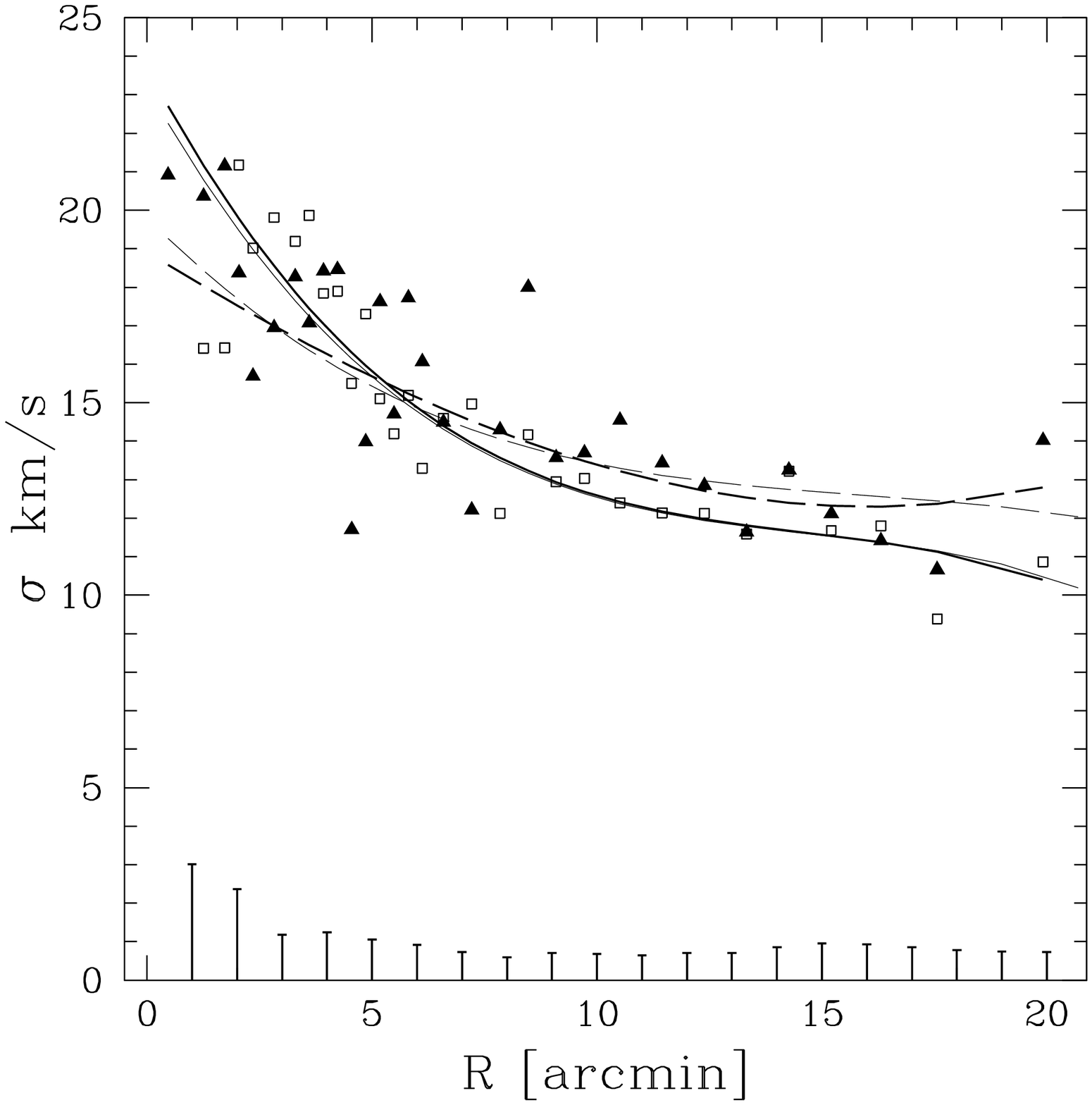}  
  \caption{Radial and tangential projected velocity dispersion as a function of 
    projected distance R for the proper motion sample used by vdV06. 
    The open squares and closed triangles  represent radial and tangential
    dispersions for vdV06, respectively. The thicker lines represent
    splines fit through the radial (solid) and tangential (dashed) velocity dispersions respectively.
    The radial and tangential proper motion velocity dispersion profiles of Sample C 
    (lighter solid and dashed respectively) fit to the data shown in Figure 1, are also included for comparison. 
    The typical size of the dispersion errors are shown at the bottom of the plot.
}
  \label{fig:dispersion}
\end{figure*}

The projected radial and tangential velocity dispersion profiles for Sample C as a function of projected distance R
is shown in Figure \ref{fig:problem} using radial bins from vL00 (Table 6).
For comparison, we show the projected radial and tangential velocity dispersion profile of the sample 
of 2295 stars used by vdV06 in Figure \ref{fig:dispersion}. In general, the velocity dispersions are very similar,
even though the evidence for a tangential anisotropy in the outer parts is weaker in our (cleaner) Sample C. 
It must also be noted that the velocity dispersions at larger radii ($>19'$) are highly uncertain,
due to the small number of stars in these radial bins.
 
\subsubsection{Radial Profile}
\label{sec:radprof}
Implicit in the methodology of Section \ref{sec:estimators} is the assumptions that the
stars with measured kinematics fairly trace the overall population that yields the density
profile, $\nu_{*}(r)$ or $\Sigma_{*}(R)$. However, crowding effects and poorer colour information 
at the centre of the cluster leads to a bias in the radial sampling of stars with
proper motions. This affects our sample. For example,  we find that fainter sub-samples of Sample C
lie at larger radii; proper motion estimates for them are precluded by crowding at smaller radii.

\begin{figure*}
  \includegraphics[width=84mm]{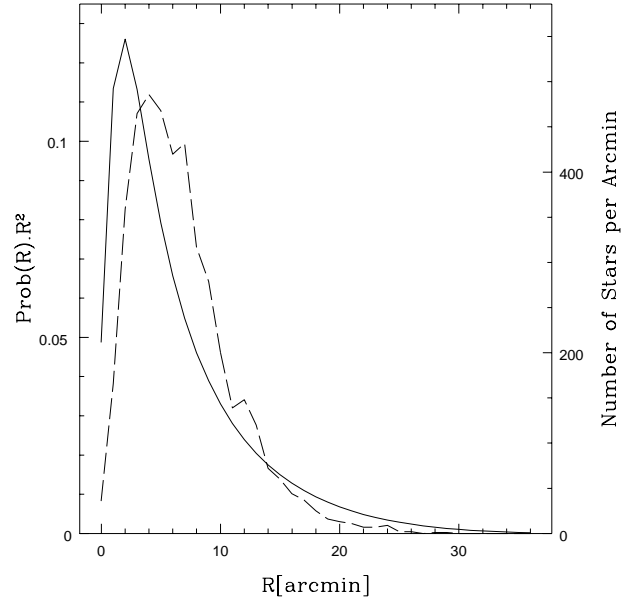}
  \caption{The probability distribution ($P(R)\cdot R^2$) for stars in radial bins of 1 arcmin
    derived from the surface density profile of Ferraro et al. (2006, solid line). 
    The dashed line represents the radial distribution of those stars in Sample C. The corresponding number 
    distribution of stars for Sample C is also given. }
   \label{fig:radprofile}
\end{figure*}

To model the surface number density, we fit a King density profile to the surface density
data (see Figure \ref{fig:problem}) published by \citet{Fer}.
The probability distribution, $P(R)$, ($\int P(R) R^{2} = 1$) for stars in each radial bin is shown in Figure \ref{fig:radprofile}. 
We can see that Sample C must be radially incomplete both near the centre of the cluster and at
larger radii. Such radially incomplete samples will affect the mass estimates from those estimators
which use an averaging process like the PME. 

For the PME, we can attempt to recover back the original density profile by repeatedly sampling the same stars.
While this is possible at the centre of the cluster because of the large number of available stars 
in the sample in those radial bins, it is difficult at the edge of the cluster ($>18'$) where the 
number of stars found in Sample C is much less than the numbers derived from the surface density profile. 
While repeatedly sampling the same stars may recover in some way the density profile, it may skew the velocity distribution. 
For example, repeatedly sampling stars that intrinsically belong to the outer edges of the cluster but which are 
projected towards the centre of the cluster, may alter the projected velocity distribution at the centre of the 
cluster. Similarly the limited number of stars at the outer ends of the cluster can bias the higher velocity moments.

\section{Mass Estimators for $\omega$ Centauri}
\label{sec:omega-mass}
In this section, we connect the methodology of Section \ref{sec:estimators}  
with the data sets of Section \ref{sec:omega-data} to estimate the mass of  $\omega$ Centauri.
We start by applying the PME (Eq. \ref{eqn:proj2}), then move on to ML-models (Section \ref{sec:MLM})
based on a King model and the Jeans equation.

\subsection{Projected Mass Estimator}
\label{sec:omega-mass-pme}
To apply the PME, we first have to adjust our Sample A and C to fairly trace the overall population 
as a function of radius (Figure \ref{fig:radprofile}). We do so by Monte-Carlo drawing members from 
Sample C so that their radial distribution matches the profile of Ferraro et al. 2006 (solid line Figure 
\ref{fig:radprofile}).

\begin{figure*}
  \includegraphics[width=84mm]{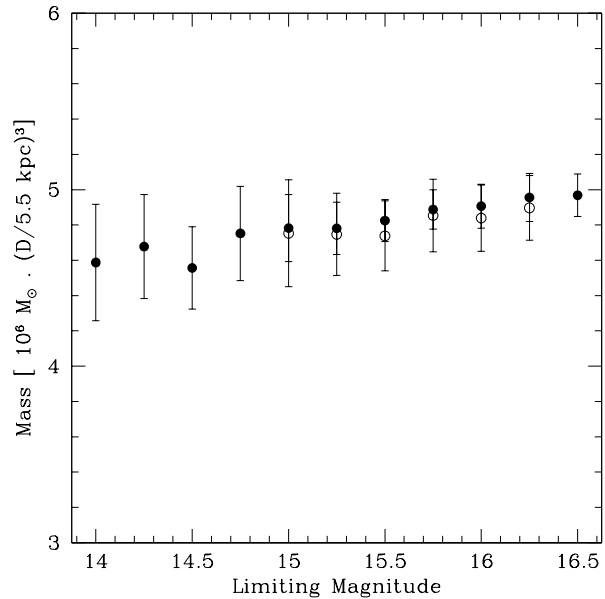}
  \caption{Results for the simple Projected Mass Estimator (Eq.
    \ref{eqn:proj2}) as a function of the limiting magnitude (B). The
    error bars ($1\sigma$) are calculated through bootstrapping. The closed symbols represent  
    Sample C stars, the open symbols represent Sample A horizontal branch stars.
    As the Projected Mass Estimator in Eq. 2 does acccount for the growing proper 
    motion errors (as a function of apparent magnitude), the mass estimate depends 
    only at the $<8\%$ level on the chosen flux limit. Note that the radial 
    distribution of the kinematic tracers depends somewhat on their magnitude, 
    as faint sources are underrepresented in the more crowded inner regions. 
    Hence some magnitude dependence may arise from the differing radial sampling.
}
 \label{fig:estimator}
\end{figure*}

The resulting PMEs are depicted in Figure \ref{fig:estimator}, with
the 1-$\sigma$ error bars calculated from bootstrapping. For bright magnitude
limits, we have only high precision measurements,
but few stars. The larger error bars demonstrate the lack of
information due to the limited amount of tracers of the system.


At magnitude 16, the PME estimates the 
mass of the cluster for Sample C as  $(4.90 \pm 0.12) \times 10^{6} M_{\odot} \left[D/5.5 \pm 0.2 kpc\right]^3 $,
while Sample A predicts $(4.84 \pm 0.19) \times 10^{6} M_{\odot} \left[D/5.5 \pm 0.2 kpc\right]^3$ for the same magnitude.

Note, that the inclusion of the error terms leads to an estimator that is virtually independent
of $M_{lim}$ (i.e., signal-to-noise). The high mass estimate reveals that the sample is radially 
incomplete especially at the center of the cluster.

\subsection{Simple Maximum Likelihood Models}
We now proceed to combine the dynamical models directly with ML estimates for the $\omega$ Centauri data
to constrain the mass of the cluster.

We start off with the classic King model \citep{K66} framework for this task, then
use the spherical isotropic Jeans equation to predict the velocity dispersion curve; in the
subsequent section we generalise this to anisotropic, axisymmetric Jeans models

\subsubsection{King Models}
\label{sec:king_models}
\begin{figure*}
   \includegraphics[width=0.33 \textwidth]{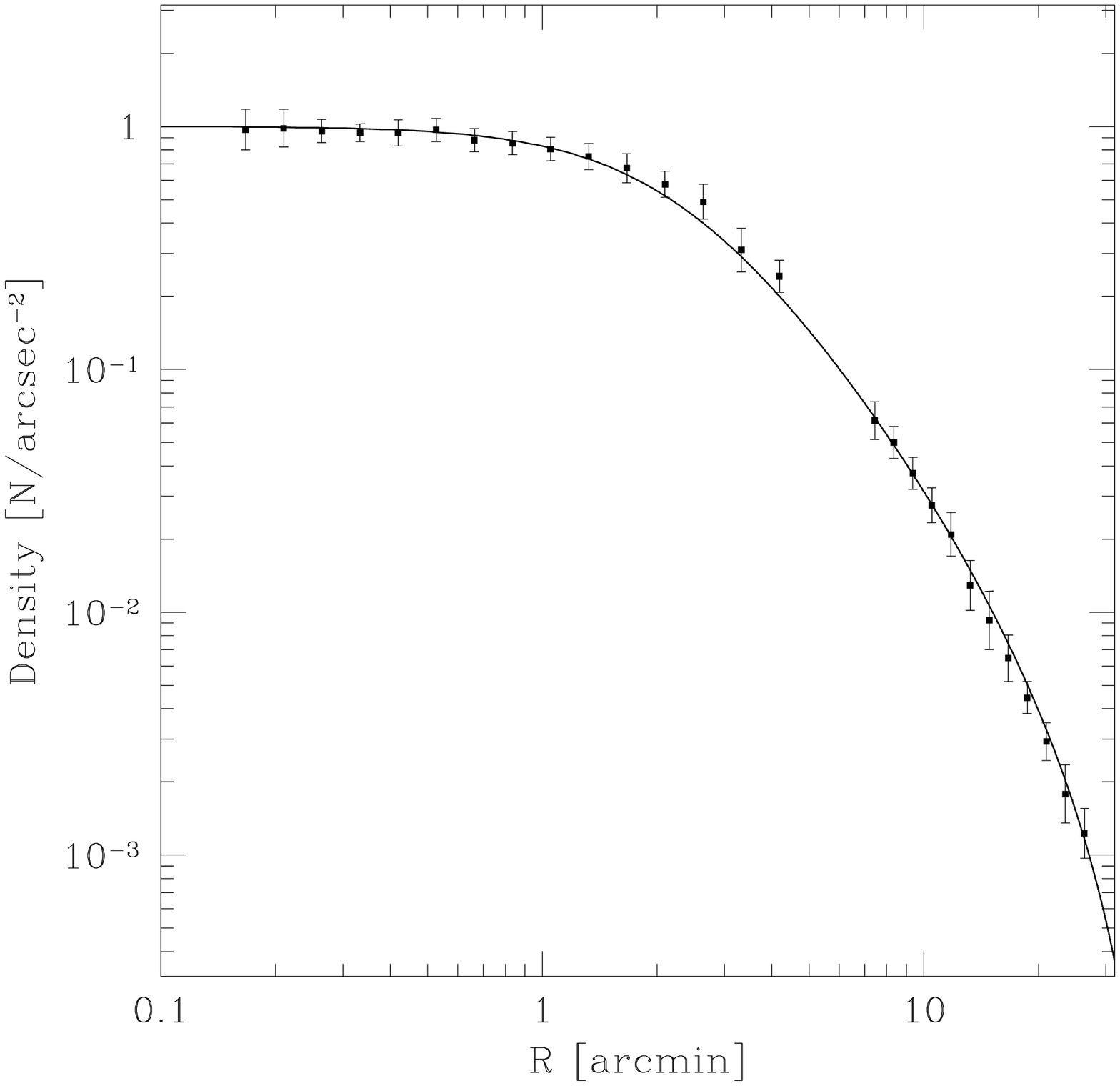}
   \includegraphics[width=0.33 \textwidth]{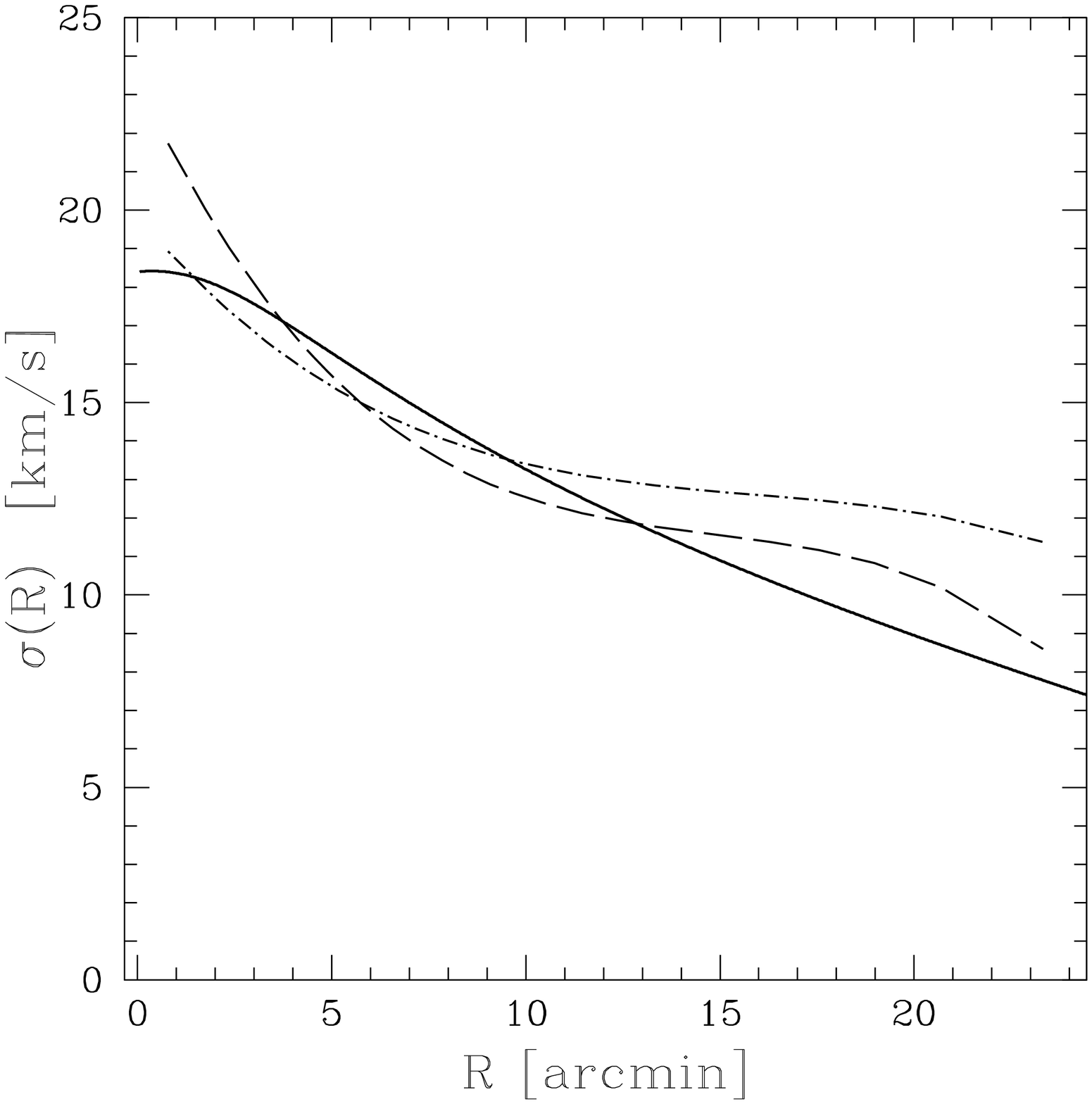}
   \includegraphics[width=0.33 \textwidth]{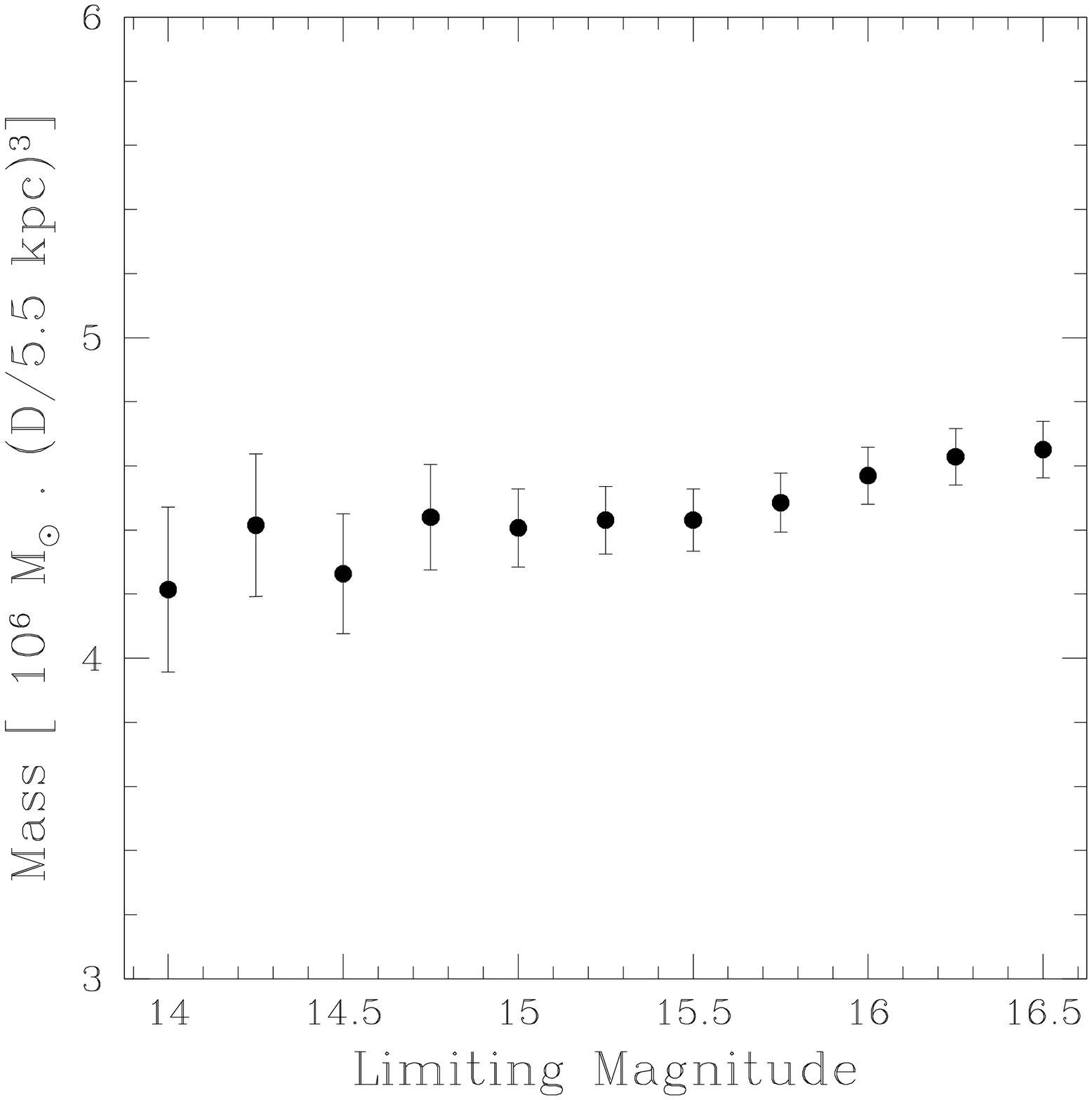}
   \caption{King model fit to $\omega$ Cen data: (a) \emph{Left}: a King profile ($R_{c}=141.676''$, $c=1.224$; King 1962, Eq. 14) 
	is fit to the number density (\citealt{Fer}) (b) \emph{Centre}: corresponding projected velocity  dispersion profile of the 
	King model as compared with the radial and tangential dispersions (dashed line and dot-dashed line respectively) of Sample 
	C stars as defined in Section \ref{sec:dataset}. (c) \emph{Right}: the resulting mass estimate of the MLM (Section \ref{sec:MLM}) 
	as a function of limiting magnitude. The right panel shows that with the proper MLM estimator, the mass estimate varies systematically by nearly $10\%$, when including fainter and fainter stars.
}
   \label{fig:single-mass-kings-model}
\end{figure*}

In Figure \ref{fig:single-mass-kings-model}, a King profile (\citealt{K62})
is fit to number density profile of $\omega$ Centauri (\citealt{Fer}) characterized
by the core radius, $r_{c}$, and the concentration, $c$. We found
$r_{c}=141.68''$ and $c=1.22$. Note that these values are different from
those originally derived from Ferraro et al. (2006) as the values they derived
correspond to a dynamic King model (Sigurdsson \& Phinney 1995).
Given the King profile, we can derive a velocity dispersion profile using 
the technique outlined in \cite{K66} or by integrating the Jeans equaton.
Even though the surface density profile provides a relatively good fit, the model
cannot reproduce the projected velocity dispersion. If we nonetheless use this King model 
and the likelihoods equation \ref{eqn:like}, we obtain
mass estimates for different magnitude limits (shown in the right panel of Figure \ref{fig:single-mass-kings-model}).

For example at $M_{lim}=16$, the mass estimate is $(4.57 \pm 0.09) \times 10^{6} M_{\odot} \left[D/5.5 \pm 0.2 kpc\right]^3$,
with the error derived using likelihood methods. To compare the various models with each other, we can define a measure of 
the scatter as the range of mass estimates at various limiting magnitudes divided by the mass estimate at $M_{lim}=16$. 
The scatter for the King model amounts to $0.096$.

\subsubsection{Spherical Isotropic Jeans Model}
The isotropic Jeans  equation for a spherical system (neglecting rotation) is:
\begin{equation}
\label{eqn:sphericaljeans}
\frac{\partial}{\partial r} (\nu_{*} \sigma^{2}) = - \nu \frac{\partial
\Phi}{\partial r}
\end{equation}
where $\sigma$ is the radial velocity dispersion, $\nu$ is the
observable number density distribution, and $\Psi$ is the
potential generated by the matter distribution at the considered
point. 

\begin{figure*}
  \includegraphics[width=0.33 \textwidth]{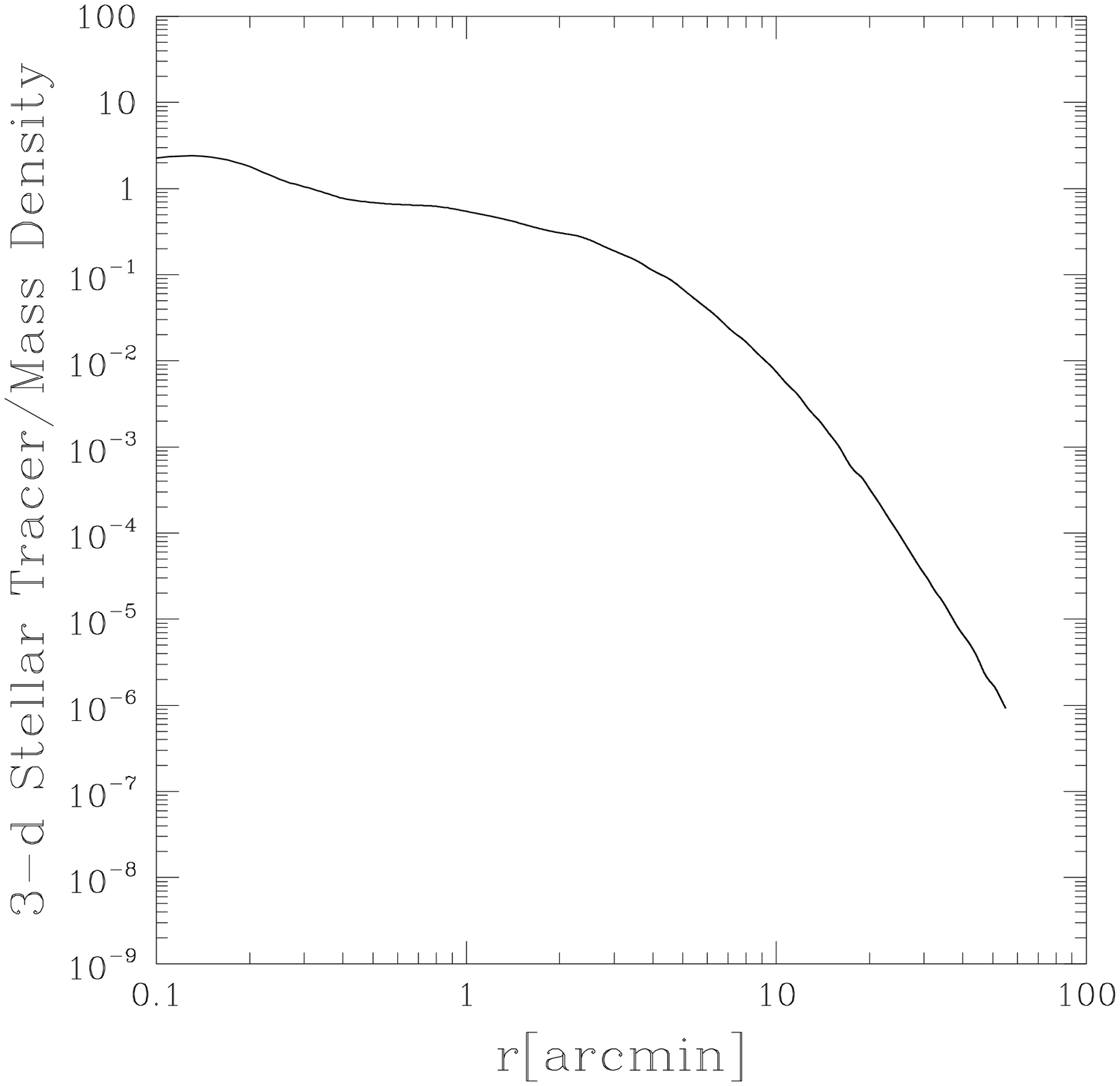}
  \includegraphics[width=0.33 \textwidth]{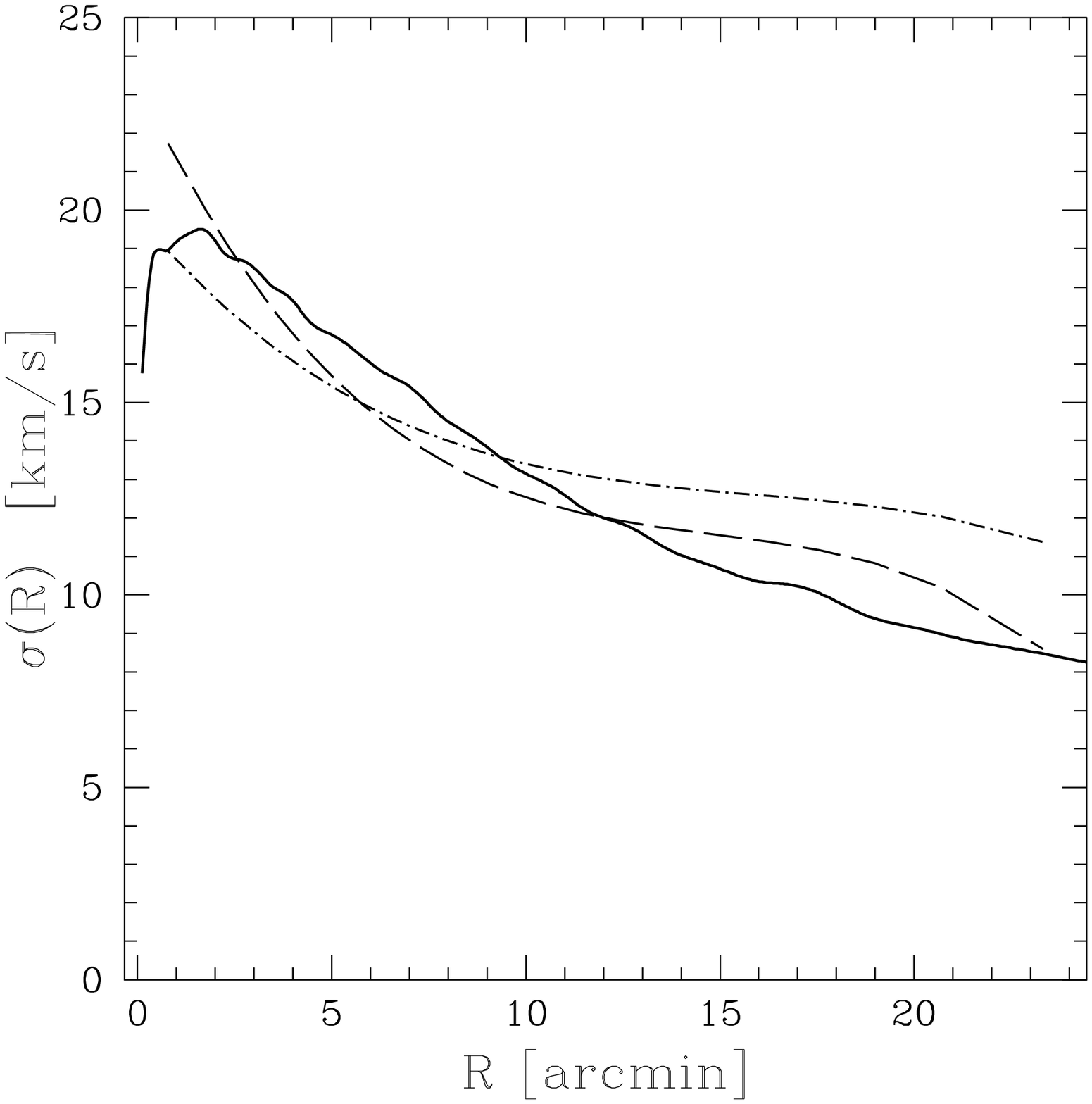}
  \includegraphics[width=0.33 \textwidth]{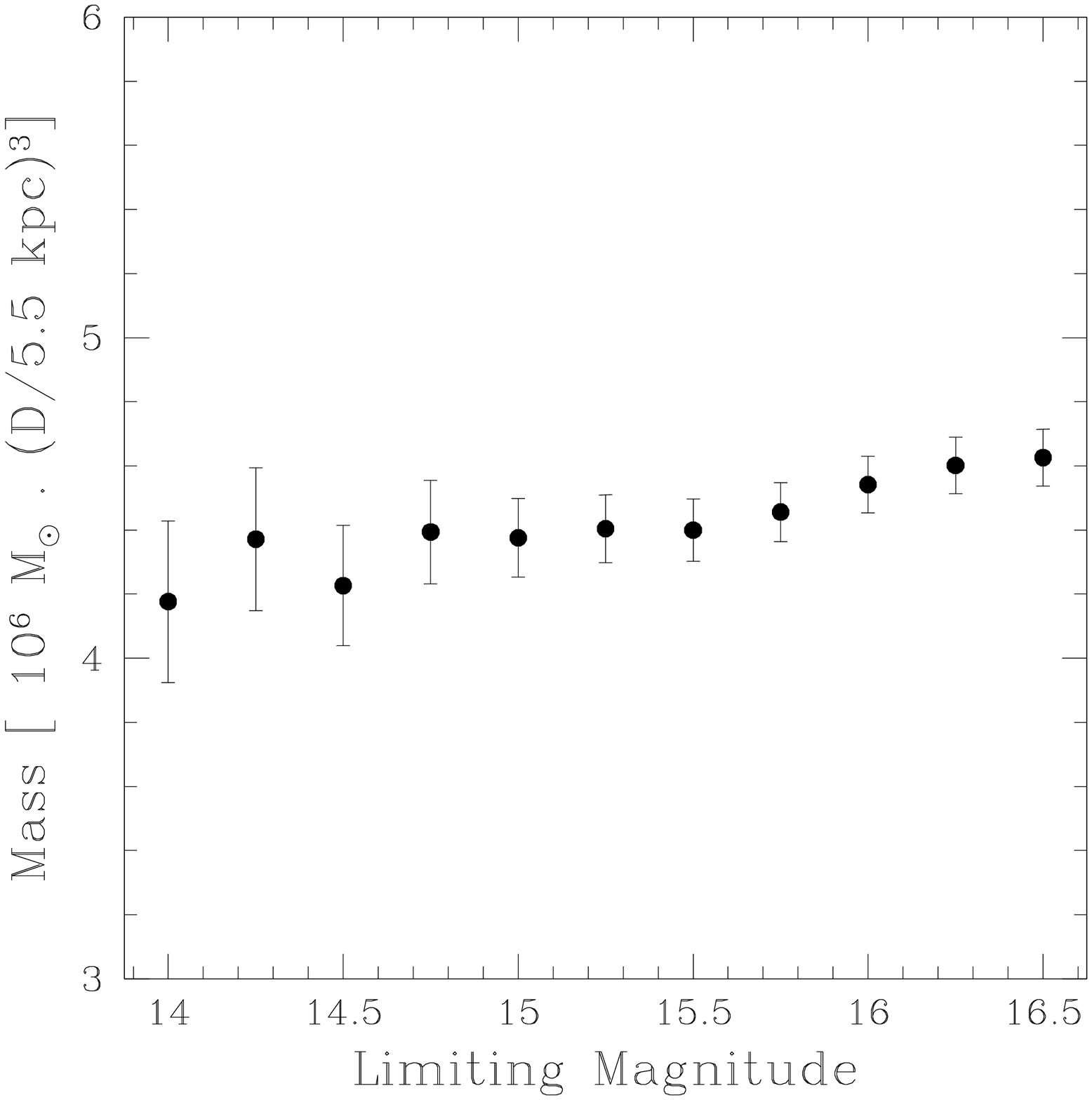}
  \caption{Spherical isotropic Jeans model fit to $\omega$ Cen data: (a) \emph{Left}: the reconstructed 3-dimensional 
	spherical number density by \citet{Mer1} using non-parametric analysis shown by the solid line. 
	(b) \emph{Centre}: the corresponding  velocity dispersion curve predicted by isotropic spherical Jeans equation (Equation \ref{eqn:sphericaljeans}) 
	is plotted along with the radial and tangential dispersion (dashed and dot-dashed line respectively) of Sample 
	C stars as defined in Section \ref{sec:dataset}. (c) \emph{Right}: the resulting mass estimate of the MLM 
	(Section \ref{sec:MLM}) as a function of limiting magnitude. We will show below that relaxing the assumption 
	of spherical symmetry, yields significantly lower mass estimates.
	}
  \label{fig:jeans-model}
\end{figure*}

We use the three dimensional number density distribution by
\citet{Mer1} derived using non-parametric analysis (left panel of Figure \ref{fig:jeans-model}). 
We assume that mass follows light, and that the density profile also defines the potential of the cluster.
The velocity dispersion is then predicted by Jeans equation. This is shown in the central panel of Figure
\ref{fig:jeans-model}. The velocity dispersion curve is then used
to predict the mass of the cluster using the MLM.

As in Section \ref{sec:king_models}, the error-bars with bright $M_{lim}$ are large due to lack of
information in the limited number of tracers. At magnitude 16, the MLM with the Jeans model estimates the
mass of the cluster as $(4.54 \pm 0.09) \times 10^{6} M_{\odot} \left[D/5.5 \pm 0.2 kpc\right]^3$
for Sample C stars. The scatter of this model as defined earlier in Section \ref{sec:king_models} is $0.099$.

The resulting MLM mass estimates of the cluster for the King model and the spherical isotropic Jeans model are
consistent with each other.

The scatter in the results is indicative of the poor fit of the spherical  models 
to the complex nature of $\omega$ Centauri. The results of the MLM are also not consistent with 
the results of the PME and those of vdV06. This indicates that
we are poorly modelling the cluster, failing to take into account the apparent rotation and the flattening of the cluster.
Hence we now turn to more sophisticated models, namely the anisotropic axisymmetric Jeans model.

\subsection{ML Estimates based on anisotropic axisymmetric Jeans Models}
\label{sec:axijeans}
Anisotropic axisymmetric Jeans modelling (e.g. \citealt{Cap}) is one way to model $\omega$ Centauri
without the restrictive approximations of isotropy and spherical symmetry. Ample evidence exists
that neither approximation holds: $\omega$ Centauri's velocity distribution is anisotropic
(\citealt{A1,A2}; vL00; vdV06). \cite{Gey} have showed that $\omega$ Centauri is one of the 
most flattened globular clusters in the Galaxy.

In the case of \cite{Cap}, the semi-isotropic axisymmetric Jeans formalism is generalised to include
anisotropy on the basis of two assumptions: a constant mass-to-light ratio $M/L$ and a constant
velocity anisotropy parameter $\beta_z=1-\overline{v_{z}^2}/\overline{v_{R}^2}$. Given a detailed
description of the surface brightness in the form of a multi-Gaussian expansion (MGE), the model can 
predict the shape of the second velocity moments ($V_{rms}$) on the basis of two free parameters 
$\beta_z$ and the inclination $i$ of the cluster. We calculate the relevant formulae for the second
velocity moments for the two components of the proper motion in Appendix \ref{app:anisotropicjeansequation}.
For the first velocity moment ($V$), additional assumptions are required. Obtaining the rotation of the cluster 
from this formalism involves the need of separating the second moments into random and streaming rotation around the symmetry axis by 
the introduction of an additional parameter (and hence assumptions) for the first order equation (in terms of 
$\kappa$; see Equation 35 of \citealt{Cap}), analogous to the Satoh (1980) approach. The advantage of using a multi-Gaussian expansion (MGE) 
is that it allows us to set the rotation ($\kappa_k$) and the anisotropy ($b_k$) individually for each Gaussian component.

\begin{figure*}
  \includegraphics[width=0.33 \textwidth]{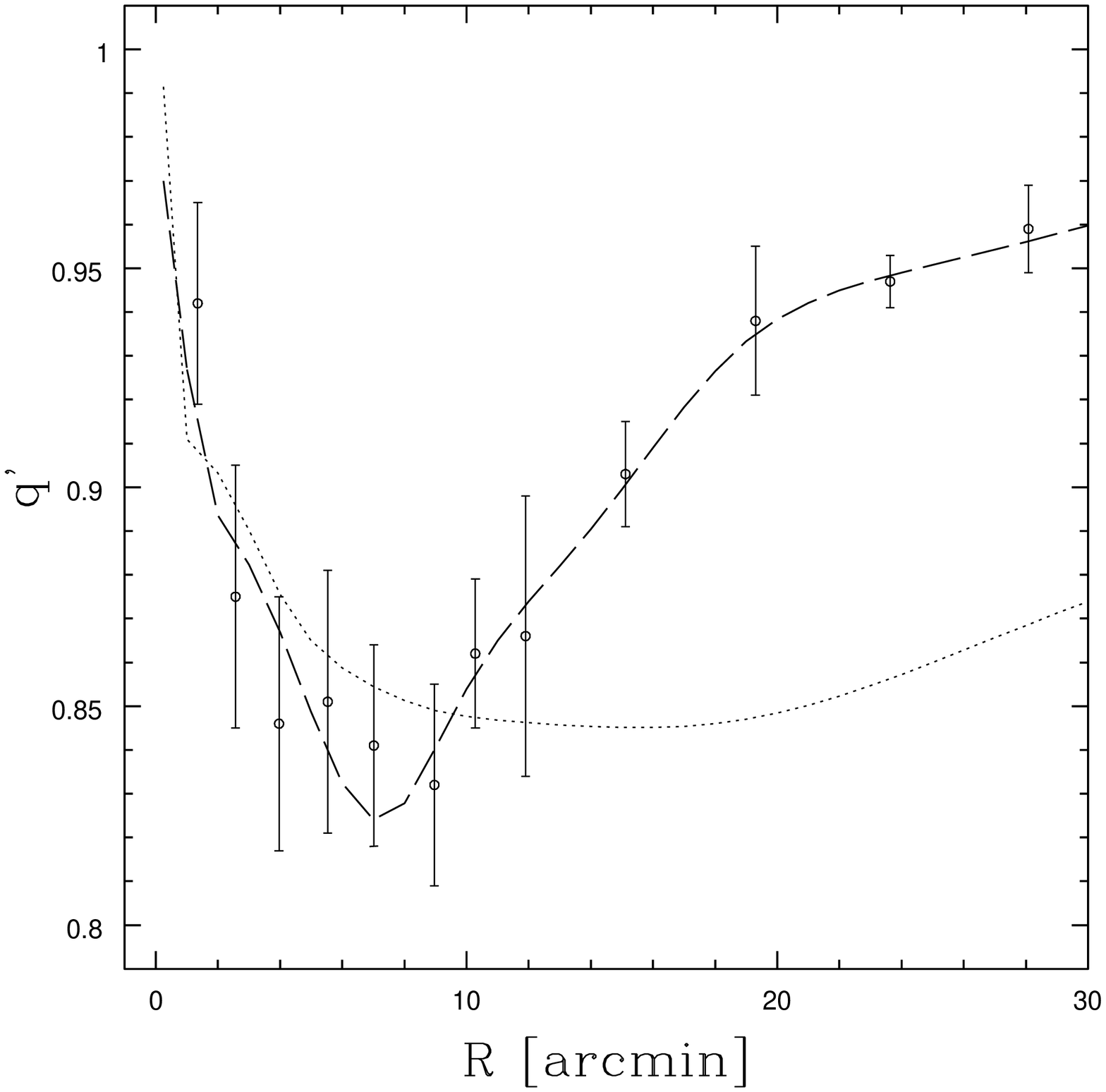}
  \includegraphics[width=0.33 \textwidth]{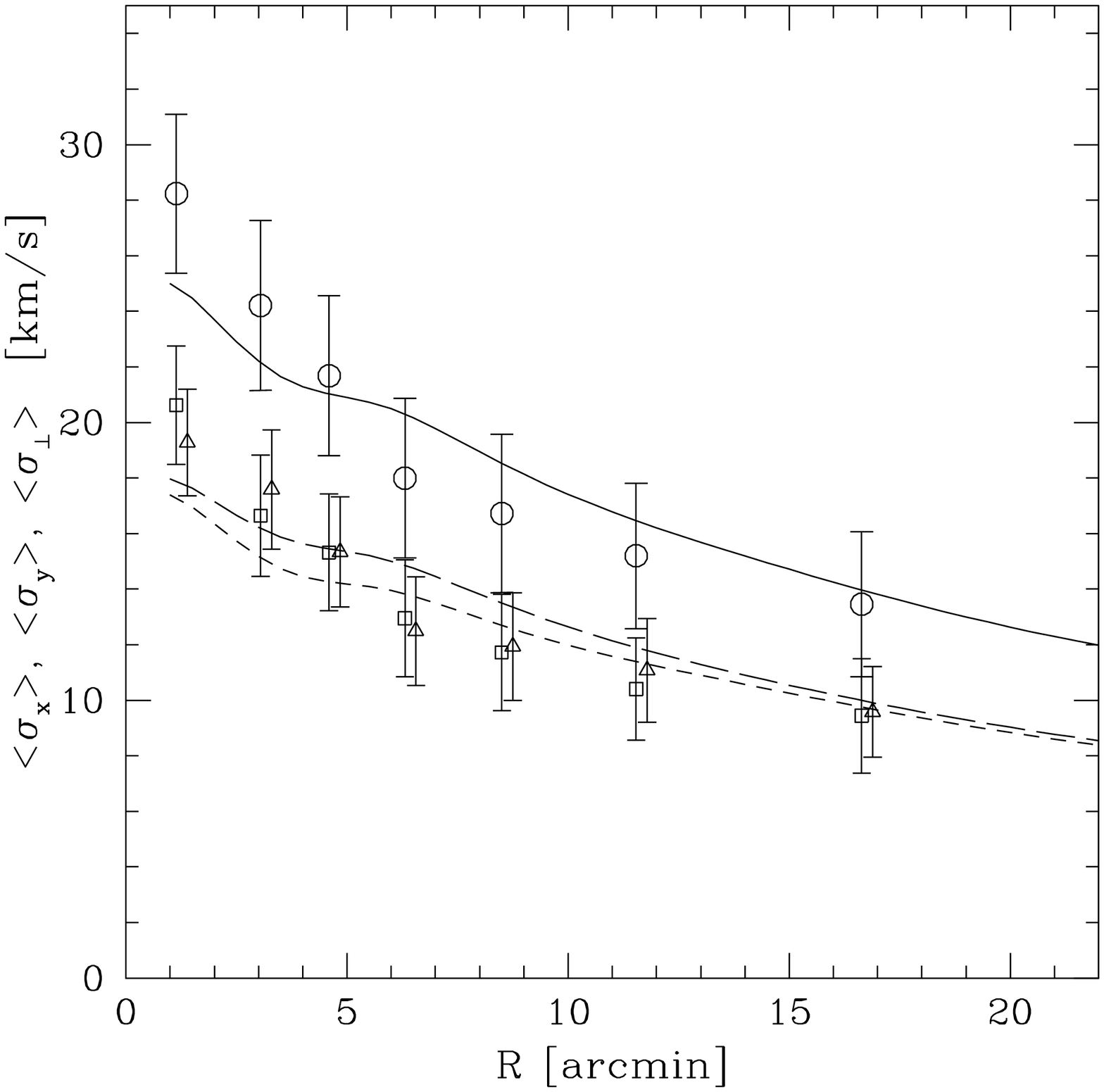}
  \includegraphics[width=0.33 \textwidth]{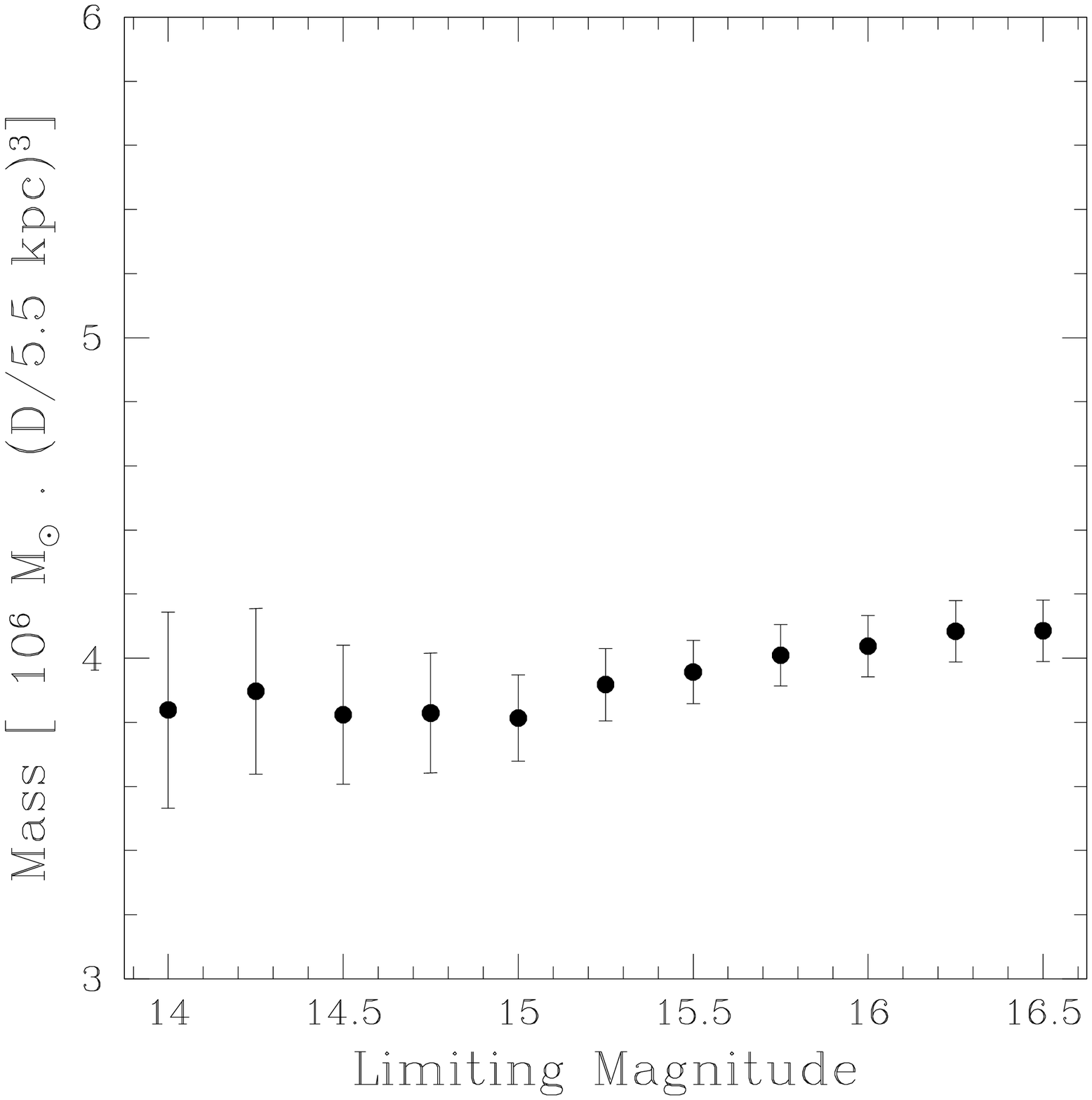}
  \caption{Anisotropic axisymmetric Jeans model: \emph{Left}: the flattening $q'$ of the projected stellar 
	density as a function of radial distance as derived by Geyer, Hopp \& Nelles (1983). The dotted 
	line shows the flattening as a function of radius for the MGE parametrization of vdV06, while the 
	long dashed line shows the flattening after adopting the parameters in Table \ref{tab:flatten}. 
	\emph{Centre}: the average velocity dispersion along the major and minor axis 
	(open squares and triangles respectively) calculated for Sample C in concentric annuli. 
	The long-dashed line and the short-dashed line depicts the dispersion along the major and minor axis 
	respectively of the best-fit anisotropic axisymmetric model. The open circles denote the total projected 
	velocity dispersion calculated in circular annuli, while the solid line is the corresponding best-fit 
	from our anisotropic axisymmetric model. \emph{Right}: the mass of the cluster as a function of 
	increasing magnitude using the anisotropic axisymmetric model. Note that the mass estimates are 
	considerably lower than in the previous Figures, where spherical symmetry was assumed.
}
\label{fig:jam_figures}
\end{figure*}

Such modelling is based on the two-dimensional surface density distribution of the stars,
which has not been published as such. However vdV06 have derived a multi-Gaussian expansion (MGE)
of the one-dimensional V-band surface brightness profile of \cite{Mey87} using 8 Gaussians.
They converted this into a two-dimensional luminosity distribution by assigning a projected
flattening $q_k'$ to each Gaussian using the data of \cite{Gey}. Unfortunately, these eight
flattened Gaussians do not result in the elliptic profile of \cite{Gey}, which increases upto $R = 7'$ 
and then falls again towards larger distances (Figure \ref{fig:jam_figures}, left panel). 
We rederived a multi-Gaussian expansion to the number density profile of \cite{Fer} once more using 8 Gaussians.
Using a Levenberg-Marquardt algorithm, we optimised the 
flattening coefficients $q_k'$ to reproduce the \cite{Gey} data as shown in Figure \ref{fig:jam_figures} (left panel).

\begin{table}
 \caption{
	The parameters of the 8 Gaussians from the MGE-fits of the number density profile of $\omega$ Cen as found by Ferraro et al. (2006).
	Similar to Table 2 of vdV06, the second column gives the central surface density 
	(in $L_{\msun} pc^{-2}$) of each Gaussian component, 
	the third column the dispersion (in arcmin) along the major axis and the fourth column the 
	projected flattening of each component which reproduces the flattening of Geyer, Hopp \& Nelles (1983). 
	The fifth column indicates the coefficient $\kappa$ used to determine the rotation of each Gaussian component.
	}
\begin{center}
 \begin{tabular}{@{}lllll}
  \hline
  k & $\Sigma_{OV}$ & $\sigma'$ & $q'$ & $\kappa$\\
  \hline
	1 & 1290.195 & 0.47557  & 1.000000  & 0.0 \\ 
	2 & 4662.587 & 1.931431 & 0.9991714 & 0.0 \\
	3 & 2637.784 & 2.513385 & 0.7799464 & 0.4 \\
	4 & 759.8591 & 3.536726 & 0.724126  & 1.1 \\
	5 & 976.0853 & 5.403728 & 0.8556435 & 0.6 \\
	6 & 195.4156 & 8.983056 & 0.9392021 & 0.0 \\
	7 & 38.40327 & 13.93625 & 0.9555874 & 0.0 \\
	8 & 8.387379 & 20.98209 & 1.0000000 & 0.0 \\
  \hline
 \end{tabular}
\end{center}
\label{tab:flatten}
\end{table}

We follow vdV06 in using an isophotal position angle of $100^{\circ}$, 
and align our coordinates axes ($x'$,$y'$) with the observed isophotal major and minor axis of $\omega$ Cen.

An important parameter in an axisymmetric model is the inclination of the cluster ($i$) which relates
intrinsic and projected flattening. This can only be constrained
by including the line-of-sight velocity data, which vdV06 have done. They arrived at a joint constraint for
the inclination and the dynamical distance under the assumption of axisymmetry: $D \tan i = 5.6 (+1.9/-1.0)$ kpc.
Using their dynamical models (an axisymmetric implementation of Schwarzschild's orbit superposition
method), they constrain further the inclination at $i=50\pm3^{\circ}$ and the 
dynamical distance. Even though \cite{krajnovic05} and \cite{Cap} have pointed out a degeneracy  
between the inclination and the anisotropy of the cluster, we adopt the inclination value from vdV06. 
The combination of the distance and the inclination we have adopted is consistent with the above
relation.

\begin{figure*}
  \includegraphics[width=0.75\textwidth]{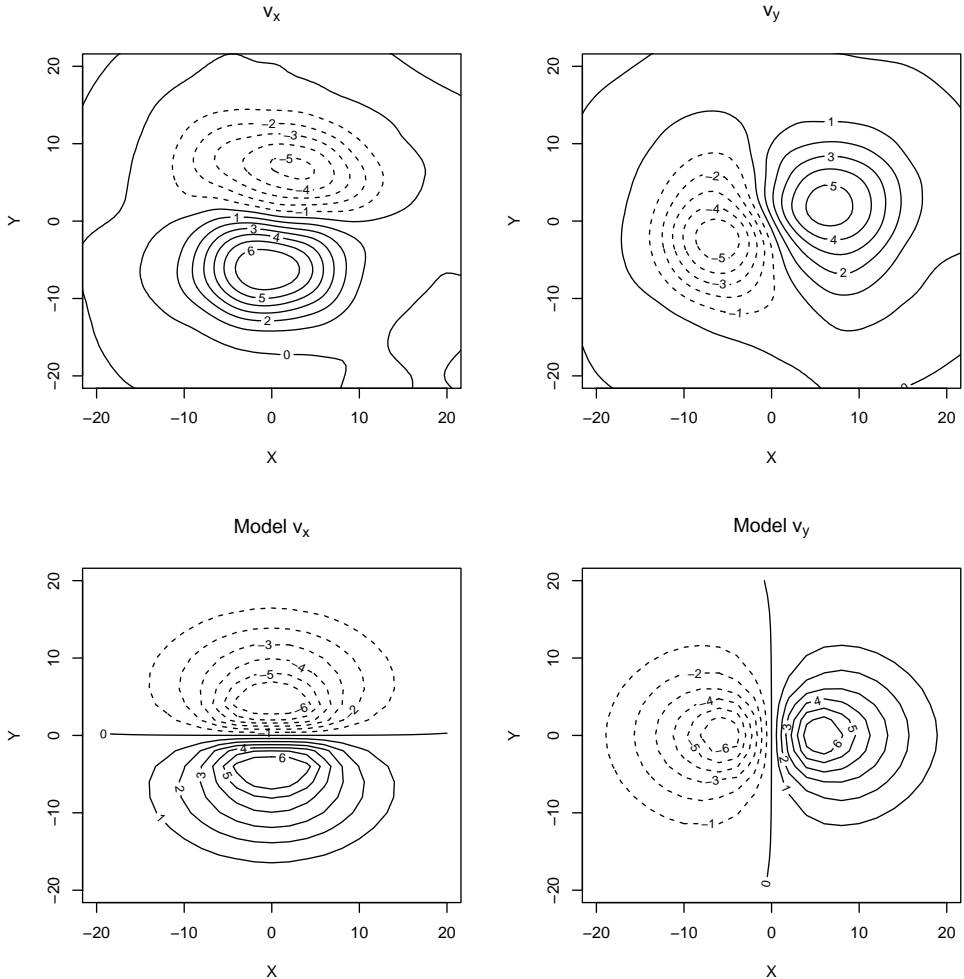}
  \caption{The top panels show the projected rotation velocity fields of $\omega$ Centauri obtained using a 
	two-dimensional spline smoothing routine, where the axes $x$ and $y$ represent the isophotal 
	major and minor axes, respectively. Velocity contours are given in the centre-of-mass frame and are spaced by 
	1 km/s, while distances are measured in arc minutes. The interpolation errors should be $<0.6$ km/s. 
	The lower two figures represent the rotation velocity fields in the $x'$ and $y'$ direction for our best-fit 
	anisotropic axisymmetric Jeans model.}
  \label{fig:rotation}
\end{figure*}

To fix $\kappa_k$ (see Equation 35 of \citealt{Cap}) required for the parametrisation of the random 
and ordered streaming motion around the cluster, we must compare the model projected differential 
rotation with those derived from the cluster data. We use the polar 
aperture grid for the proper motions of vdV06. To derive the differential rotation of the cluster 
we use the non-parametric approach of \cite{Mer1} on the binned data, where we 
minimise the ``penalised log likelihood'' (see Equation 4 of \citealt{Mer1})
and used a `thin-plate smoothing spline' routine found in the R-package `FIELDS'. 
We forced the velocities at large radii ($>19'$) to go down to zero, so that the main features 
at lower radii do not get swamped by the thin-plate spline routine. The differential
rotation velocity contours are shown in the upper panel of Figure \ref{fig:rotation}.
Note that any solid-body rotation of a cluster cannot be determined 
from the differential proper motions.  

Given the inclination ($i$), we can proceed to derive $\kappa_k$ for each Gaussian component by comparing 
the rotation curves with those predicted by the model. The resulting values of $\kappa_k$ are given in Table 
\ref{tab:flatten}. It must be noted that the peaks of rotation do not 
lie along the isophotal major axis along the y direction. If $\omega$ Centauri were exactly axisymmetric, the 
kinematic minor axis would coincide with the isophotal minor axis. However, Figure \ref{fig:rotation} shows 
that this is only approximately true. A similar feature was noted by \cite{Mer1} for $\omega$ Centauri 
while deriving the rotation curves from the line-of-sight velocities.

The anisotropic Jeans model allows us to set the anisotropy of each Gaussian component ($b_k$). However we 
follow the practical approach of setting a global anisotropy parameter $b$ (see Section \ref{sec:effects}), 
which we can derive from the proper motion data after fixing the rotation of the cluster. Similar to Equation 
\ref{eqn:rotation}, we can construct a maximum likelihood estimator. The probability for measuring a given 
individual proper motion $\mu_x$ and $\mu_y$, given the parameters $M$ and $b$ are:
\begin{eqnarray}
p_{model}( \mu_x | X, Y, M, b ) & = & \frac{1}{\sqrt{2 \pi\ \sigma^{2}(M, X, Y, b)}} \nonumber \\
& & \exp \left\{ \frac{- (\mu_x - \overline{v}_x(M, X, Y, b)) ^{2}}{2\ \sigma^{2}(M, X, Y, b)} \right\} 
\label{eqn:anisotropy}
\end{eqnarray}
with a corresponding equation for $\mu_y$. To obtain the global anisotropy parameter $b$, 
we marginalise over the mass $M$. We then maximise the likelihood for our Sample C, varying only the parameter b. 
Following standard practice, we can parameterise this anisotropy with the variable 
$\beta_z\equiv1-\overline{v_{z}^2}/\overline{v_{R}^2}=1-1/b$, where $z$ is the line-of-sight direction. 
Using Sample C, if the inclination of the cluster is fixed at $i=50\pm3^{\circ}$, we find $\beta_z = 0.03 \pm 0.04$ 
indicating that overall the cluster is close to isotropy.

\begin{figure*}
  \includegraphics[width=0.8 \textwidth]{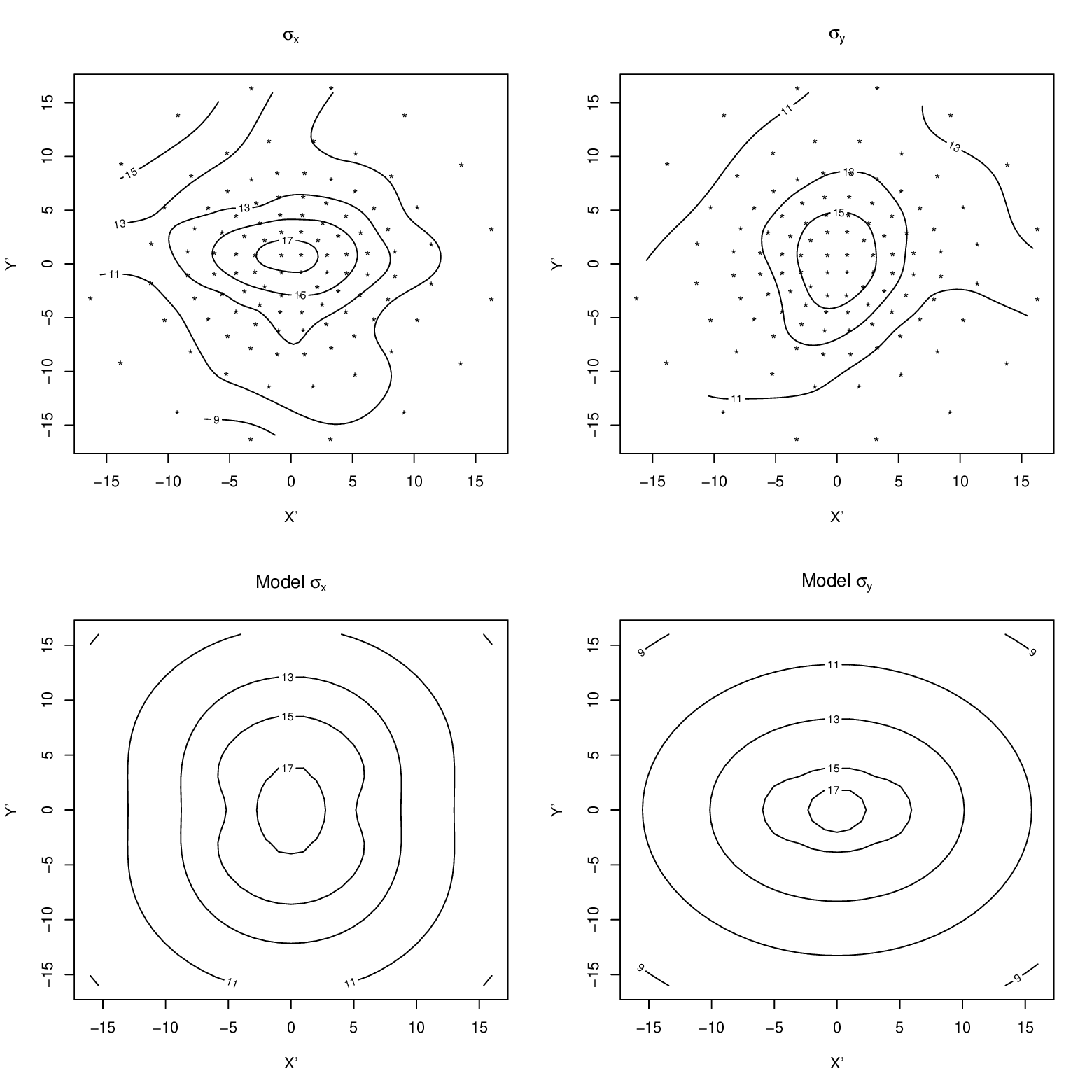}
  \caption{\emph{Top}: the total projected velocity dispersions along the photometric major and minor axis respectively calculated 
    using a thin-plate smoothing spline (see Equation 11 of Merritt et al. 1997). The dispersion was calculated 
    using the polar apertures given in Table 3 of vdV06 and then smoothened with a thin-plate spline routine. 
    The centres of the polar apertures are indicated by a dot. \emph{Bottom}: the velocity dispersions 
    as predicted by the best-fit anisotropic axisymmetric Jeans model. 
  }
  \label{fig:2D-dispersion}
\end{figure*}

We also calculate the 2-dimensional velocity dispersion profile of Sample C along 
the major and minor axis using a thin-plate spline (see Equation 11 of Merritt et al. 1997) and compare them
with the model velocity dispersion profiles (see Equations \ref{eq:second_moment_x} and \ref{eq:second_moment_y}). 
The results are shown in Figure \ref{fig:2D-dispersion}. 
The dispersions were calculated using the polar apertures given in Table 3 
of vdV06 and the standard techniques outlined in Appendix A and then fed to the thin-spline routine. 
Note that the elongation of the iso-dispersion velocity contours
along the minor axis is not consistent with the assumptions of axisymmetry.

With an estimate of the global anisotropy parameter $\beta_z$, we can now obtain the mass of the cluster 
using the MLM (Equation \ref{eqn:like}) in combination with Equation \ref{eqn:anisotropy} for each component 
of the proper motion using Sample C. For each star in Sample C, we calculate the first and the second velocity moments
for both the components of the proper motion (see Equations \ref{eq:first_moments_projection_final_x}, 
\ref{eq:first_moments_projection_final_y}, \ref{eq:second_moment_x} and \ref{eq:second_moment_y}). 
The final mass estimate using our own derived values of flattening in Table \ref{tab:flatten}, 
is shown in Figure \ref{fig:jam_figures}. The good fit of the model is reflected in the low scatter in the mass 
estimates: the scatter of this model as defined in \ref{sec:king_models} is $0.76$.
The mass of the cluster using the anisotropic Jeans model is $(4.05 \pm 0.10) \times 
10^{6} M_{\odot}\left[D/5.5 \pm 0.2 kpc\right]^{3}$ at the limiting magnitude $B=16.0$.

\subsubsection{Effects of Anisotropy, Inclination, Rotation and Flattening on the Mass Estimate}
\label{sec:effects}
Within such a modelling context, it would be instructive to separate the effects due to inclination,
rotation, anisotropy and flattening of the cluster. 

While the value of the inclination of the cluster is consistent with the distance (5.5 kpc) we have adopted 
and the relation $D \tan i = 5.6 (+1.9/-1.0)$ kpc, we do have a range of acceptable inclination $ i = 45.5 (+8.5/-5.9)$ degrees. 
However, with the assumptions of axisymmetry, the observed projected flattening of the cluster sets further constraints on 
the inclination of the cluster. From Eq. 14 of \cite{Cap}, the inclination is constrained by the flattening: $i > \cos^{-1}(q'_{k}) \,\,\, \forall k$.
The fourth Gaussian of the MGE in Table \ref{tab:flatten} sets the lower limit of the inclination at $i=43.6^{\circ}$. 
At $i=45.5^{\circ}$, the mass inferred from Sample C is $(3.90 \pm 0.10) \times 10^{6} M_{\odot} \left[D/5.5 \pm 0.2 kpc\right]^{3}$. 
At the lower end $i=44^{\circ}$, the mass is $(3.82 \pm 0.10) \times 10^{6} M_{\odot} \left[D/5.5 \pm 0.2 kpc\right]^{3}$, while at 
the higher end of $i=54^{\circ}$, the mass is $(4.15 \pm 0.10) \times 10^{6} M_{\odot} \left[D/5.5 \pm 0.2 kpc\right]^{3}$. 
Hence, the inclination of the cluster affects the mass estimate upto $6\%$. 

By neglecting the cluster rotation in the model, the mass inferred based on Sample C 
is $(4.22 \pm 0.07) \times 10^{6} M_{\odot} \left[D/5.5 \pm 0.2 kpc\right]^{3}$; if we include rotation, we get 
$(4.05 \pm 0.10) \times 10^{6} M_{\odot} \left[D/5.5 \pm 0.2 kpc\right]^{3}$. Hence accounting for rotation affects
the mass estimate for the cluster by only $4.2\%$.

To isolate the effect due to anisotropy, we calculate the `best' mass of the cluster for
$\beta_z=0$ and $\beta_z=0.1$. We get the same mass estimate in both cases. Hence we conclude that in
such a modelling environment, the anisotropy of the cluster does not contribute much to the mass of the cluster.
This is because the observed anisotropy is very small.

\begin{figure*}
  \includegraphics[width=0.48 \textwidth]{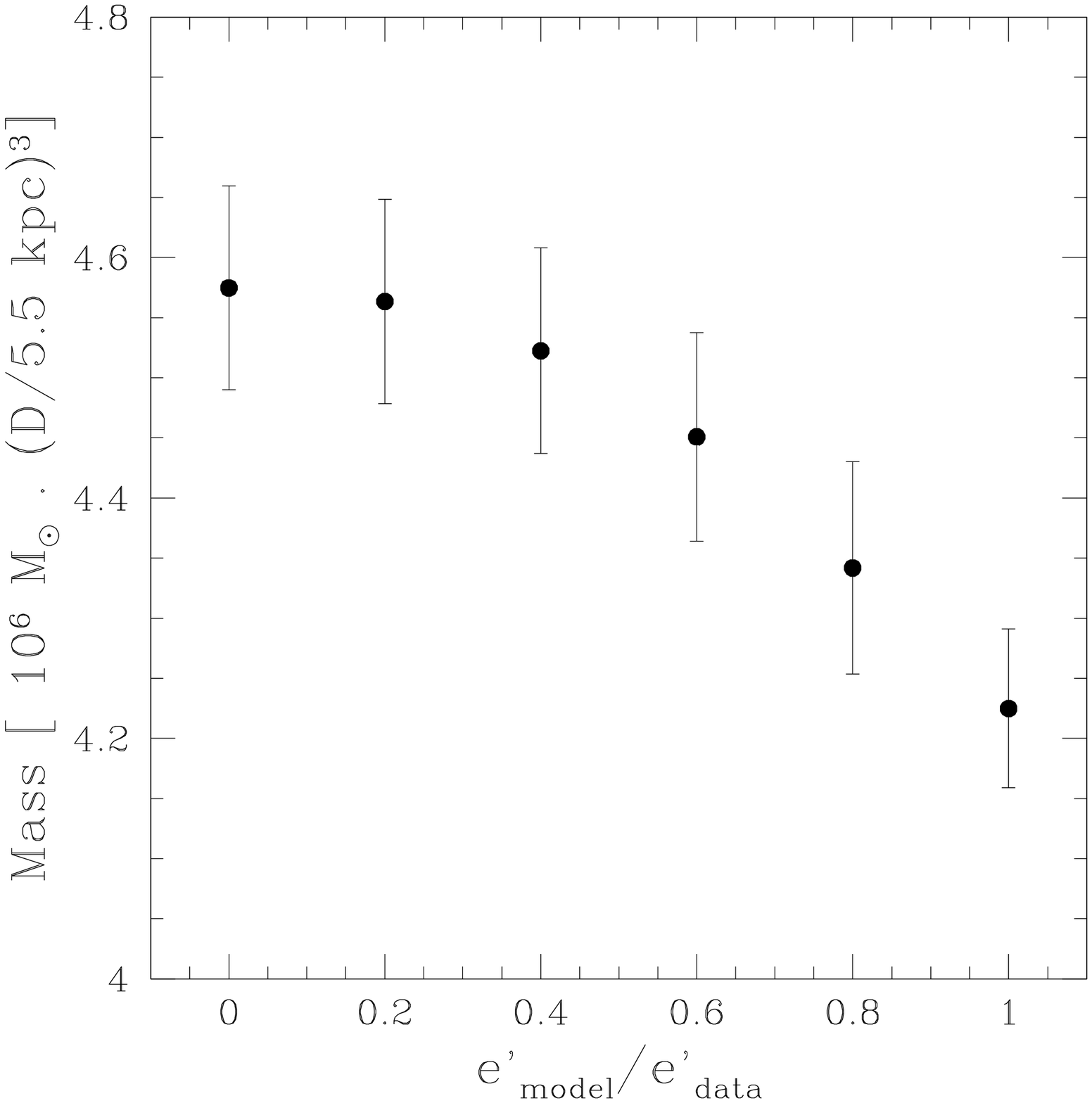}
  \caption{Cluster flattening and mass estimates: using the flattening profile of the 8 Gaussians in Table \ref{tab:flatten}, 
    we study the effect of the variation in the estimated mass of the cluster as a function of  increasing flattening. 
    We do not consider the rotation of the cluster. We define the projected eccentricity 
    $e_{j}' = (1 - q'^{2}_{j})^{0.5}$ for each Gaussian component. The left end of the x-axis indicates 
    the absence of flattening, while the right end indicates a flattening which replicates the projected 
    flattening profile in Table \ref{tab:flatten}. 
    The figure shows the estimated mass of the cluster using the complete Sample C. This shows the effect of 
    modelling flattening on the estimated mass of the cluster. Notice that the results reduce back to the 
    simple isotropic spherical case in the absence of flattening (see Figure \ref{fig:jeans-model}). If we were to 
    include rotation to the modelling process, the mass of the cluster  is estimated to $(4.05 \pm 0.10) 
    \times 10^{6} M_{\odot} \left[D/5.5 \pm 0.2 kpc\right]^{3}$ (not shown in the figure).}
  \label{fig:qresults}
\end{figure*}

Accounting for the flattening of the cluster, however, is important. This is already apparent from the fact
that the masses inferred via the flattened models are $10\%$ lower than our earlier estimates
based on spherical models. To explore this further, we define for each Gaussian component a 
projected eccentricity $e_{j}' = (1 - q'^{2}_{j})^{0.5}$. 
We then study how the estimate of the mass of the cluster varies as a function of the flattening parameter of the model in terms of the relative projected eccentricity $e'_{model}/e'_{data}$, where $0$ corresponds to no 
flattening, while $1$ would correspond to the full flattening derived in Table 1. This results in a difference of 
nearly $8.3\%$ in the mass estimates as shown in Figure \ref{fig:qresults}, which 
indicates the importance of properly determining the fattening of the cluster to determine its mass. 
Including also the effects of rotation along with the flattening profile of Table \ref{tab:flatten} further 
brings the mass to $(4.05 \pm 0.10) \times 10^{6} M_{\odot} \left[D/5.5 \pm 0.2 kpc\right]^{3}$ which is 
consistent with the results of vdV06. 

At the dynamical distance of 4.8 kpc adopted by vdV06, this corresponds to a mass estimate of 
$(2.69 \pm 0.07) \times 10^{6} M_{\odot} \left[D/4.8 kpc\right]^{3}$

\subsection{Discussion of $\omega$ Centauri Mass Estimates}
\label{sec:dis}
Our analysis presented here has made it clear that the major discrepancy in the estimated masses of $\omega$ Centauri arises primarily 
due to the choice of the modelling process. By taking into account both the flattening of the cluster 
and its inclination using axisymmetric models, and to a lesser extent by accounting for the rotation, 
we can derive consistently a mass estimate that has only $\sim3\%$ error for a given distance. 

However, even in our most complex modelling scenario, which included both anisotropy and axisymmetry, 
we had to draw on vdV06 to get a cluster inclination. Yet at the same time, we observed significant deviations in the 
2-d velocity dispersion map. Our modelling has also shown that compared to the flattening, the inclination, 
anisotropy and the rotation, have less impact on the mass estimates. More accurate
constraints must be put on the flattening of the cluster through more accurate two dimensional surface density 
profiles. Further, $\omega$ Centauri with its complex formation history is an ideal candidate for more sophisticated 
Schwarzschild modelling as attempted by vdV06. A perhaps more immediate step would be to incorporate radial velocity 
data in the MLM approach established here.

Irrespective of the modelling context, the largest uncertainty in our mass estimates based on proper motion data, 
arises from the adopted cluster distance, which scales the estimated mass as $\sim \left[D/5.5 \pm 0.2 kpc\right]^{3}$. 
In this paper, we have assume a distance of 5.5 kpc derived from RR Lyrae stars \citep{Delprincipe}, with 
random and systematic distance errors of only $3\%$ each. This estimate is consistent
with the estimate of the eclipsing binary OGLEGC 17 \citep{Thompson}, but substantially different from the 
dynamical distance of 4.8 kpc estimated by vdV06. Reliable dynamical distances depend on a host of factors - 
but crucially on a proper estimation of the velocity dispersions, which in turn is dependent on the proper 
selection of the sample. The estimate of vdV06  which is lower than that of \cite{Delprincipe}, could have 
come about if there is an over-estimation of the proper motion dispersion due to unrecognized interlopers
or an under-estimation of the line-of-sight velocity dispersion. Intrinsic to these 
dispersion estimates is also a proper correction for perspective rotation (which is distance dependent) and 
solid-body rotation. \cite{Bon} report a systematic offset between reported absolute photometric distances 
and dynamical distances in the clusters $\omega$ Centauri and 47 Tuc, with dynamical distance estimates being lower.
Therefore, relying on the precise and completely independent distance estimate from \cite{Delprincipe}
appears preferable. The $3.8\%$ distance errors translates into a $\approx 12\%$ error in the mass estimate. 

As this paper is geared as a proof-of-concept paper of what is possible with proper motion data, 
we have steered away from directly using the line-of-sight velocity data. However, tackling the
issues mentioned above warrants the use of the same. In such cases however, more sophisticated
modelling schemes than what we have considered in this paper would be required to determine the 
dynamical distance of the cluster, which we postpone to a later paper.

\section{Conclusions}
\label{sec:concl}
In this paper, we implement two independent methods to determine the mass of a 
cluster from the proper motion data. Both the Projected Mass Estimator (PME) and the 
Maximum Likelihood modelling (MLM) provide unbiased and robust estimates of the mass of
the cluster. However the use of the PME is limited due to  a number of practical difficulties in
sampling the cluster. For the MLM, we use an analytic King model, a spherical isotropic Jeans 
equation model and an axisymmetric, anisotropic Jeans equation model.

We apply these approaches to the extensive ground-based proper-motion data-set of $\omega$ Centauri 
by vL00. Using the high precision photometry of \cite{Rey2}, we construct a kinematically unbiased 
member samples, that should have less contamination than the samples used in previous analyses.

Among the three models used in the MLM context, the anisotropic axisymmetric Jeans equation with some rotation 
is by far the best model accounting for the photometric and kinematic data of $\omega$ Centauri, though
there exists compelling evidence of deviations away from axisymmetry. Neglecting the flattening and 
inclination of the cluster by using spherical models overestimates the 
mass of the cluster by $10\%$. This accounts also for the large spread in reported masses of  $\omega$ Centauri.

The mass of the cluster is $(4.05 \pm 0.10) \times 10^{6} M_{\odot} \left[D/5.5 \pm 0.2 kpc\right]^{3}$. 
The largest uncertainty in the absolute mass estimate is the distance to $\omega$ 
Centauri which enters as the third power.  With a distance error of $3.8\%$ this amounts to a mass error of $\sim 12\%$.

While our results are mostly consistent with those of vdV06, the difference arises due to our adoption of the 
new RR lyrae distance estimates and due to our more accurate reconstruction of the flattening and the shape of the MGE 
parametrization leading to an increase in the mass estimate by $7\%$.

This estimate has a $2.5\%$ statistical error, but $12\%$ uncertainty arising from the distance errors.

The authors thank Glenn van de Ven for helpful comments on the manuscript. The authors also thank the referee for her
patience and for helpful advice.

\appendix
\section{Velocity Dispersion}
\label{app:veldisp}
The projected velocity dispersion $\sigma_{\perp}$ of the cluster is needed for the kinematic removal of outliers, and can be calculated in the context of a maximum-likelihood analysis (vdV06). For a given cluster member star with velocity $v_{i}$ at a given projected distance $R$, the probability that the cluster has a dispersion $\sigma(R)$ is given by
\begin{equation}
p_{i}(v_{i} | \, \sigma(R)) = \int p_{obs}(v_{i}| \, v , \delta v_{i}) \, p_{dist} (v | \, \sigma(R) )\, \mathrm{d}v,
\end{equation}
where $p_{obs}$ is the probability of measuring the velocity $v_i$ as defined in Equation \ref{eqn:pobs} and $p_{dist}$ is the probability of a star in the cluster to have a certain proper motion velocity $v$ given a velocity distribution parameterized by the velocity dispersion $\sigma(R)$.

The projected velocity dispersion $\sigma(R)$ is then derived by maximising the likelihood function 
for a given number of stars drawn from a concentric ring centred around a projected distance $R$: 
\begin{equation}
\mathcal{L}(\forall  v_{i} |\, \sigma(R)) = \sum_{i=1}^{N}\ln p_{i}(v_{i} | \, \sigma(R)),
\end{equation}
with the  $1\sigma$ uncertainty resulting from $\mathcal{L} = 0.5$. Usually, Gaussians are a good low-order approximation for the velocity distribution $p_{dist}(v)$, thus simplifying the process.

We can also modify the distribution function to account for the probability of the star to belong to 
the cluster (first introduced by \citealt{Prada}; \citealt{W1}):
\begin{equation}
p_{dist}(v) = \alpha p_v + [1 - \alpha]\frac{1}{2 v_{max}}
\end{equation}
where $\alpha$ is the probability of the star to belong to the cluster, $p_v$ is the regular Gaussian 
distribution and $v_{max}$ is the maximum projected velocity in the particular annular ring where 
the dispersion is being calculated.

\section{Anisotropic Jeans Equation}
\label{app:anisotropicjeansequation}
Following the formalism of \cite{Cap}, we can calculate the second moments and the mean velocity along the two perpendicular axes projected on the plane of the sky. \cite{Cap} have already indicated that the projected proper motion dispersion can be written via single quadratures. Similar to equation 28 of \cite{Cap}, we can derive the analogue expressions for the proper motions using the formulas in Appendix A of \cite{evans94}. The $y'$ component of the velocity dispersion is:
\begin{eqnarray}\label{eq:second_moment_y}
    \lefteqn{\Sigma\,\overline{v_{\rm y'}^2}(x',y') = 4\pi^{3/2} G \int_0^1 \sum_{k=1}^N \sum_{j=1}^M\, \nu_{0k}\, q_j\, \rho_{0j}\, u^2} \nonumber \\
  & & \times\frac{
  \sigma_k^2 q_k^2 \left(\sin^2 i + b_k \cos^2 i\right) + \mathcal{D}\, x'^2\cos^2 i
  }{
  {\left(1-\mathcal{C} u^2\right) \sqrt{\left(\mathcal{A} + \mathcal{B}\cos^2 i\right)
       \left[1-(1-q_j^2)u^2\right]}}
  } \nonumber \\
  & & \times\exp\left\{-\mathcal{A}\left[x'^2+\frac{(\mathcal{A}+\mathcal{B}) y'^2}{\mathcal{A} + \mathcal{B}\cos^2 i}\right]\right\}\dd u,
\end{eqnarray}
where we retain the definitions of  $\nu_{0k}$, $\mathcal{A}$, $\mathcal{B}$, $\mathcal{C}$ and  $\mathcal{D}$. For the sake of completeness we repeat these definitions once again.

\begin{equation}\label{eq:a}
    \mathcal{A}=\frac{1}{2}\left(\frac{u^2}{\sigma_j^2} + \frac{1}{\sigma_k^2}\right)
\end{equation}
\begin{equation}\label{eq:b}
    \mathcal{B}=\frac{1}{2}\left\{\frac{1-q_k^2}{\sigma_k^2 q_k^2}
    +\frac{(1-q_j^2)u^4}{\sigma_j^2\left[1-(1-q_j^2)u^2\right]}\right\}
\end{equation}
\begin{equation}
    \mathcal{C}=1-q_j^2-\frac{\sigma_k^2\, q_k^2}{\sigma_j^2}
\end{equation}
\begin{equation}
    \mathcal{D}=1-b_k\, q_k^2 - \left[(1-b_k)\,\mathcal{C} + (1-q_j^2)\, b_k\right] u^2.
\end{equation}

Similarly, the $x'$ component of the velocity dispersion is:
\begin{eqnarray}\label{eq:second_moment_x}
    \lefteqn{\Sigma\,\overline{v_{\rm x'}^2}(x',y') = 4\pi^{3/2} G \int_0^1 \sum_{k=1}^N \sum_{j=1}^M\, \nu_{0k}\, q_j\, \rho_{0j}\, u^2} \nonumber \\
  & & \times\frac{
  b_k \sigma_k^2 q_k^2 + \mathcal{D}\frac{(\mathcal{A} + \mathcal{B})^2 \cos^2 i}{(\mathcal{A} + \mathcal{B}\cos^2 i)^2}\, y'^2 + \frac{\mathcal{D}\sin^2 i}{2(\mathcal{A} + \mathcal{B}\cos^2 i)}
  }{
  {\left(1-\mathcal{C} u^2\right) \sqrt{\left(\mathcal{A} + \mathcal{B}\cos^2 i\right)
       \left[1-(1-q_j^2)u^2\right]}}
  } \nonumber \\
  & & \times\exp\left\{-\mathcal{A}\left[x'^2+\frac{(\mathcal{A}+\mathcal{B}) y'^2}{\mathcal{A} + \mathcal{B}\cos^2 i}\right]\right\}\dd u
\end{eqnarray}

We can also obtain the proper motion first velocity moments of the whole MGE model similar to equation 39 of \cite{Cap}. The MGE model uses an anisotropic analogue of Satoh (1980) approach, as seen in equation 37 of \cite{Cap}, where $\kappa$ fixes the rotation for each Gaussian component.
\begin{equation}\label{eq:satoh}
    \nu\overline{v_\phi} = \left[\nu\sum_{k=1}^N \kappa_k^2 \left([\nu\overline{v_{\phi}^2}]_k - [\nu\overline{v_R^2}]_k\right)\right]^{1/2}. \nonumber
\end{equation}

\begin{eqnarray}\label{eq:first_moments_projection_final_y}
    \lefteqn{\Sigma\,\overline{v_{\rm y'}}(x',y') =
    2\sqrt{\pi G}\; x'\cos i} \\
    & & \times  \int_{-\infty}^{\infty}
    \left[\nu\int_0^1\!\! \sum_{k=1}^N \sum_{j=1}^M
    \frac{\kappa_k^2 \nu_k q_j \rho_{0j} \mathcal{H}_j(u)\, u^2 \mathcal{D}}
    {1-\mathcal{C} u^2}\dd u \right]^{1/2}\!\!\!\! \dd z'.\nonumber
\end{eqnarray}

\begin{eqnarray}\label{eq:first_moments_projection_final_x}
    \lefteqn{\Sigma\,\overline{v_{\rm x'}}(x',y') =
    - 2\sqrt{\pi G}\; y'\cos i} \\
    & & \times  \int_{-\infty}^{\infty}
    \left[\nu\int_0^1\!\! \sum_{k=1}^N \sum_{j=1}^M
    \frac{\kappa_k^2 \nu_k q_j \rho_{0j} \mathcal{H}_j(u)\, u^2 \mathcal{D}}
    {1-\mathcal{C} u^2}\dd u \right]^{1/2}\!\!\!\! \dd z' \nonumber \\
    & & + 2\sqrt{\pi G}\; \sin i \nonumber \\
  & & \times  \int_{-\infty}^{\infty}
    z' \left[\nu\int_0^1\!\! \sum_{k=1}^N \sum_{j=1}^M
    \frac{\kappa_k^2 \nu_k q_j \rho_{0j} \mathcal{H}_j(u)\, u^2 \mathcal{D}}
    {1-\mathcal{C} u^2}\dd u \right]^{1/2}\!\!\!\! \dd z'. \nonumber
\end{eqnarray}

The velocity dispersion can then be calculated as:
\begin{eqnarray}
\sigma_{x'}^2 = \overline{v_{x'}^2} - \overline{v_{x'}}^2 \\
\sigma_{y'}^2 = \overline{v_{y'}^2} - \overline{v_{y'}}^2 
\end{eqnarray}

We used the adaptive quadrature routines found in QUADPACK. Both the integrands in equation \ref{eq:first_moments_projection_final_y} and \ref{eq:first_moments_projection_final_x} can be ill-defined for certain combinations of anisotropy and rotation coefficients $\kappa_k$ towards the centre of the cluster.

\label{lastpage}
\end{document}